\documentclass[]{aastex62}
\usepackage{color}
\usepackage{epsfig}
\usepackage{graphicx}
\usepackage{amsmath}
\usepackage[english]{babel}
\usepackage[titletoc,title]{appendix}
\usepackage{txfonts}
\usepackage{times}
\usepackage{natbib}

\begin{document} 
\title{Galaxy Formation in Sterile Neutrino Dark Matter Models} 
\author{N. Menci$^1$, A. Grazian$^1$, A. Lamastra$^{1,2}$, F. Calura$^3$, M. Castellano$^1$, P. Santini$^1$}
\affil{$^1$INAF - Osservatorio Astronomico di Roma, via Frascati 33, I-00078 Monteporzio, Italy}
\affil{$^2$Space Science Data Center - ASI, via del Politecnico SNC, I-00133 Roma, Italy}
\affil{$^3$INAF - Osservatorio Astronomico di Bologna, Via Gobetti 93/3, 40129 Bologna, Italy}

\begin{abstract}
We investigate galaxy formation in models with dark matter (DM) constituted by sterile neutrinos. Given their large parameter space, defined by the combinations of sterile neutrino mass $m_{\nu}$ and mixing parameter $\sin^2(2\theta)$ with active neutrinos, we focus on models with  $m_{\nu}=7$ keV, consistent with the tentative 3.5 keV line detected in several  X-ray spectra of clusters and galaxies. We consider i) two resonant production models with $\sin^2(2\theta)=5\,10^{-11}$ and $\sin^2(2\theta)=2\,10^{-10}$, to cover the range of mixing parameter consistent with the 3.5 keV line; ii) two scalar-decay models, 
 representative of the two possible cases characterizing such a scenario: a freeze-in and a freeze-out case. We also consider thermal Warm Dark Matter with particle mass $m_X=3$ keV. Using  a semi-analytic model, we compare the predictions for the different DM  scenarios with a wide set of observables. We find that comparing the predicted evolution of the stellar mass function, the abundance of satellites of Milky Way-like galaxies, and the global star formation history of galaxies with observations does not allow to disentangle the effects of the baryonic physics from those related to the different DM models. On the other hand, the distribution of  the stellar-to-halo mass ratios, the abundance of faint galaxies in the UV luminosity function at $z\gtrsim 6$, and the  specific star formation and age distribution of local, low-mass galaxies constitute potential probes for the considered DM scenarios. We discuss how next observations with upcoming facilities will enable to rule out or to strongly support DM models based on sterile neutrinos.
\end{abstract}

\shorttitle{Galaxy Formation in Sterile Neutrino Dark Matter Models}
\shortauthors{N. Menci et al.}
\keywords{cosmology: dark matter -- galaxies: formation}
%
%________________________________________________________________

\section{Introduction}
Dark matter (DM) is the main driver of galaxy  formation. Its nature defines the power spectrum of density fluctuations which collapse to form cosmic structures, thus determining the abundance and the properties of galaxies at the different mass scales. E.g.,  the standard 
 Cold Dark Matter (CDM) scenario is based on candidates constituted by massive ($m_X>0.1$~GeV)  thermal relics (WIMPS) or  condensates of light axions (with mass $\sim 10^{-5}-10^{-1}$~eV), characterized by thermal velocities small enough to  make  density perturbations gravitationally unstable down to mass scales negligible for galaxy formation.  
Correspondingly, the r.m.s. amplitude of density fluctuations continues to increase for decreasing mass scales, yielding an ever increasing abundance of 
dwarf galaxies down to sub-galactic mass scales ($M\approx 10^7-10^9$ $M_{\odot}$). 
%, and steep inner density profiles $\rho\sim r^{\gamma}$ with $\gamma\lesssim -1$ in all galactic DM halos.  
 However, several observations concerning the dwarf galaxy population seem to challenge such a scenario (see Bullock \& Boylan-Kolchin 2017 for a recent review and for complete account of the existing literature). 
%First,  
E.g., the predicted abundance of low-mass DM halos is much larger than the observed abundance of dwarf galaxies. The issue is most acute for satellite 
galaxies  (Klypin et al. 1999; Diemand et al. 2008; Moore et al. 1999 Springel et al. 2008), but is present also in the field (see, e.g., Zavala et al. 2009; Schneider et al. 2017). 
%Second, the observed density of dwarf galaxy cores is lower than predicted by CDM simulation (McGaugh, Rubin, de Blok 2001; Marchesini et al. 2002; Simon et al. 2005; de Blok et al. 2008; Kuzio de Naray, McGaugh, de Blok 2008). 
A possible solution can be sought in the effects of feedback from supernovae (e.g. Larson 1974) and from UV background (e.g. Efstathiou 1992), which can suppress or even prevent (see, e.g., Sawala et al. 2016) star formation in low-mass galaxies, thus strongly reducing the number of luminous 
galaxies in CDM models, bringing them in closer agreement with observations (see, e.g. Garrison-Kimmel et al. 2017). 
%In addition, repeated feedback events 
%can progressively yield an expansion of the DM density profiles in dwarf galaxies leading to the observed cored profiles at low redshifts $z\lesssim 1.5$(see, e.g., Pontzen, Governato 2012; Madau, Shen, Governato 2014). 
However, such  a suppression of the baryon-to-DM content of galaxies 
can potentially lead to a tension with the observed values.  In fact, the kinematics of dwarfs galaxies (with stellar masses $M_*=10^{6}-10^{8}\,M_{\odot}$) indicate that they are hosted by DM  halos with mass smaller than predicted by CDM models (Garrison-Kimmel et al. 2014; Ferrero et al. 2012; Papastergis 2015; Papastergis \& Shankar 2016). 

While the possibility that feedback effects can provide  a simultaneous solution to all the above issues within the CDM framework is still matter of debate,  an increasing attention is being devoted to alternative DM models with suppressed power spectra at small galactic scales $M\approx 10^{8}-10^{9}$ $M_{\odot}$ with respect to the CDM case. This is also motivated by the fact that both direct (Aprile et al. 2012, 2016; 
 Akerib et al. 2014) and indirect (see, e.g., Adriani et al. 2013, Ackermann et al. 2015) DM detection experiments have failed to provide a definite confirmation of the CDM  scenario. Also, no evidence for CDM candidates with mass $10^2-10^4$~GeV has been found in experiments at the Large Hadron Collider  (see, e.g., Ade et al. 2016), while experiments aimed to detect axions as DM components have produced no evidence in the explored portion of the parameter space (Graham et al. 2015, Marsh et al. 2016). 
 
 The combination of astrophysical issues with the lack of detection of CDM candidates has motivated several groups to investigate galaxy formation in a number of alternative models. 
Among the proposed DM candidates, a prominent class is constituted by models that assume DM to be constituted by lighter particles with mass 
$m_X$ in the keV range (see de Vega \& Sanchez 2010). The simpler assumption is to consider such particles to be thermal relics (Warm Dark Matter, WDM, see Bode, Ostriker \& Turok 2001) 
resulting from the freeze-out of particles initially in thermal equilibrium in the early Universe (like, e.g., gravitinos, see Steffen 2006 for a review). 
Their larger thermal velocities (corresponding to larger free-streaming lengths) suppress structure formation at scales $M=10^7$--$10^9$ $M_\odot$, depending only on the value of $m_X$ (since a thermalized species has no memory of the details of its production). 
Such a  one-to-one correspondence between the WDM particle mass and the suppression in the power spectrum at small scales has allowed to derive limits on $m_X$ by comparing the predictions from $N$-body WDM simulations or semi-analytic models with the abundance of observed ultra-faint galaxies. On this basis, different authors have derived limits ranging from $m_X\geq 1.5$~keV (Lovell et al. 2014) to $m_X\geq 1.8$ keV (Horiuchi et al. 2014), $m_X\geq 2$~keV (Kennedy et al. 2014) and $m_X\geq 2.3$~keV (Polisensky, Ricotti 2011) from the abundance of 
local dwarf satellites, while at higher redshifts, $z\approx 6$, a limit $m_X\gtrsim 1.5$~keV has been derived from the UV luminosity functions of faint galaxies down to $M_{\rm UV}\approx -16$ (Schultz et al. 2014; Corasaniti et al. 2017). The tighter constraints achieved so far $m_X\geq 3.3$~keV (at 2-$\sigma$ c.l.) for WDM thermal relics are derived by comparing small scale structure in the Lyman-$\alpha$ forest of high- resolution ($z > 4$) quasar spectra with hydrodynamical $N$-body simulations (Viel et al. 2013). While all the above methods are affected by uncertainties related to baryon physics, a baryon-independent limit $m_X\geq 2.5$ (2-$\sigma$ c.l.) keV has been derived by Menci et al. (2016) from the abundance of $z=6$ galaxies observed by Livermore et al. (2017; see also Bouwens et al. 2017a). The overall result from the studies above is that a 
limit $m_X\gtrsim 2.5-3$ keV for thermal WDM candidates constitute a rather robust indication.

An alternative possibility for keV-scale DM  is constituted by models based on sterile neutrinos (SN hereafter). In such scenarios the power spectra 
 are non-thermal, and depend not only on the assumed mass of the SN $m_{\nu}$, but also on the the production mechanism. In particular, 
 for SN produced from oscillations of active neutrinos, the power spectrum depends also on the mixing angle $\theta$ defining the 
 admixtures $\sin^2(2\theta)$ with the active neutrinos. While earlier models (Dodelson \& Widrow 1994) required relatively large mixing angles to produce the observed abundance of DM, more recent scenarios assume an enhancement due to resonant production in the presence of a non-vanishing lepton asymmetry 
 (Shi \& Fuller 1999), thus allowing for extremely small mixing angles $\sin^2(2\theta)\leq 10^{-9}$. 
These models received a particular interest in the literature in recent years (see Adhikari et al. 2017 for a complete review). This is due to both solid fundamental physics motivations (right-handed neutrinos constitute a natural extension of the Standard Model to provide mass terms for active neutrinos, see Merle 2013) and to the fact that such particles constitute the simplest candidates (see, e.g., Abazajian et al. 2014) for a DM  interpretation of the  potential X-ray line at energy $E\approx 3.5$ keV in stacked observations of galaxy clusters and in the Perseus cluster with the Chandra observatory (Bulbul et al. 2014); independent indications of a consistent line in XMM-Newton observations of M31 and the Perseus Cluster (Boyarsky et al. 2014) have been 
 followed by measurements in different objects (the Galactic center and  other individual clusters, Iakubovskyi et al. 2016; measurements of the cosmic X-ray Background towards the COSMOS Legacy and CDFS survey fields, Cappelluti et al. 2017) and  
 from other observatories like {\it Suzaku}  (Urban et al. 2015; Franse et al. 2016) and NuSTAR 
 (observations of the COSMOS and Extended Chandra Deep Field South survey fields,  Nerenov et al. 2016).
In fact, the tiny admixtures $\sin^2(2\theta)$ with the active neutrinos allow the decay of SNs, resulting into photon emission at energies close to $m_{\nu}/2$ with emission $F\propto \sin^2(2\theta)$. The non-detection of such a line in several systems (see, e.g., Malyshev et al. 2014; Anderson et al. 2015; Bulbul et al. 2016; Ruchayskiy et al. 2016;  Adhikari et al. 2017 for an extended discussion)  yields effective \emph{upper} limits on the mixing angle as a function $m_{\nu}$.  
E.g., for $m_{\nu}=7$ keV (the value consistent with the tentative 3.5 keV line) non-detections yield a limit $log \sin^2(2\theta)\lesssim -9.7$;  
 while still consistent with the range of values corresponding to the tentative 3.5 keV line ($-10.6\leq log \sin^2(2\theta) \leq -9.5$), such a limit 
 from non-detections is effective in  ruling out sterile neutrino models based on the non-resonant production mechanism by Dodelson \& Widrow (1994). Thus, the  present observational situation leaves open the possibility for sterile neutrino models 
based on resonant production by oscillations with active neutrinos (RP models, Shi \& Fuller 1999), and to models in which SNs are produced by 
the decay of a scalar particle (SD), presented in detail in Merle \& Totzauer (2015 and references therein). 

Given the interest in the above DM models with suppressed power spectra, several studies of galaxy formation have been carried out in WDM scenarios  using either semi-analytic models (Menci et al. 2012, 2013; Benson et al. 2013; Nierenberg et al. 2013; Kang et al. 2013; Dayal et al. 2015; Bose et al. 2017), or hydrodynamical simulations (Herpich et al. 2014; Maio \& Viel 2015; Lovell et al. 2016; Wang et al. 2017), addressing   the global galaxy properties (like luminosity and stellar mass functions and galaxy star formation) and the properties of satellite galaxies around Milky Way-like galaxies in WDM scenarios with $m_X$ ranging from 1 to 3 keV. However, a global exploration of galaxy formation in sterile neutrino DM scenarios is still missing, although the investigation of specific issues has been undertaken using high-resolution simulations (Bose 2017; Lovell et al. 2016; 2017a,b). In this paper, we tackle with this task by exploring the impact of assuming different existing sterile neutrino DM models on the observable properties of galaxies, including stellar mass and luminosity functions, satellite abundances, $L/M$ ratios and star formation properties.  To this aim, we use a state-of-the-art semi-analytic model (SAM; see Somerville \& Dav\'e 2015 for a review) with different initial power spectra, 
each corresponding to a selected sterile neutrino model. Given the large parameter space of such DM models, determined by the possible combination of sterile neutrino mass $m_{\nu}$ and mixing parameter $\sin^2(2\theta)$, we choose to focus this work on models with fixed sterile neutrino mass 
$m_{\nu}=7$ keV, i.e., on models which can be consistent with the tentative 3.5 keV line in the X-ray spectra of clusters and galaxies discussed above. In particular, we consider i) two RP models of SN with mixing angles $\sin^2(2\theta)=5\,10^{-11}$ and $\sin^2(2\theta)=2\,10^{-10}$, 
to cover the range of mixing parameter consistent with the tentative 3.5 keV line; ii) two SD models 
 representative of the two possible cases characterizing such a scenario: a freeze-in, and a freeze-out case case (see sect. 2 for further details). For comparison, we also show our prediction for the CDM case and for thermal WDM with mass $m_X=3$ keV. 

The paper is organized as follows. In Sect. 2 we describe the model set up: sect. 2.1 provides a brief description of the SN models we consider, in sect. 2.2  we describe how we implement such DM  models in our SAM, while in sect. 2.3 we  recall how the baryonic processes affecting galaxy formation are described in the SAM, 
and we present our strategy to fix the model free parameters. 
 In Sect. 3 we present our results concerning: the local properties of DM halos (Sect. 3.1, satellite abundance, stellar-to-halo mass ratios), the evolution of the galaxy population (sect 3.2, evolution of the stellar mass and luminosity distributions), and the star formation (3.3,  specific star formation, star formation histories, ages of stellar populations, colors). The aim is  to  investigate to what extent the comparison with the different  observables can help to disentangle the effects of baryon physics (in particular of 
 feedback) from the specific effects of the different assumed DM models. Sect 4 is devoted to discussion and conclusions. 
 
 Throughout the paper round 'concordance cosmology' values have been assumed for the cosmological parameters: Hubble constant $h=0.7$ in units of 100 km/s/Mpc,  total matter density parameter $\Omega_0=0.3$ and baryon density parameter $\Omega_b=0.045$.
  
\section{The Model}
\subsection{Dark Matter Scenarios}

The  evolution of the DM condensations on spatial scales $r$ is determined by the  power spectrum $P(k)$ of DM perturbations (in terms of the wave-number $k=2\pi/r$) 
that can be computed from the momentum distribution function of the DM. As discussed in the Introduction, 
we consider five DM models besides CDM, chosen as to yield appreciable suppression in the power spectrum with respect to CDM 
 but still consistent (albeit marginally) with existing bounds from X-ray observations and structure formation; thus the corresponding power spectra 
 (shown in fig. 1) constitute a representative set of different possible forms for such borderline cases. In detail, we consider the following DM models  : \newline 
 \begin{itemize}
\item CDM: in this case we adopt the power spectrum $P_{CDM} (k)$ given in Bardeen et al. (1986). 
\item thermal WDM with particle mass $m_X=3$ keV; in this case the suppression of power spectrum $P(k)$ with respect to the CDM case is 
${P(k)/ P_{CDM}(k)}=[1+(\alpha\,k)^{2\,\mu}]^{-10/\mu}$ with $\alpha=0.049$ $
[{\Omega_X/ 0.25}]^{0.11}$ $[{m_X/  {\rm keV}}]^{-1.11}\,[{h/0.7}]^{1.22}\,h^{-1}\,{\rm Mpc}$ and  $\mu=1.12$ (Bode, Ostriker \& Turok 2001; see also Viel et al. 2006; Destri, de Vega, Sanchez 2013). As noted in the Introduction, for thermal WDM the spectrum depends solely on the assumed 
 particle mass $m_X$. For our assumed mass $m_X=3$ keV the WDM spectrum is suppressed by 1/2 with respect to CDM at the 
half-mode mass scale $M_{1/2}= 4\cdot 10^8\,h^{-1}\,M_{\odot}$, with the suppression rapidly increasing at smaller masses (see fig. 1). 
\item RP1 model: Resonantly produced sterile neutrinos with $m_{\nu}=7$ keV and  $\sin^2(2\theta)=2\,10^{-10}$. 
Since  for each combination of $m_{\nu}$ and $\sin^2(2\theta)$ the lepton asymmetry is fixed to the value required to yield the right DM abundance, 
our choice corresponds to a lepton asymmetry $L_6\approx 8$ (in units of $10^{-6}$, see, e.g., Boyarsky et al. 2009;  Abazajian 2014).
The momentum distribution  strongly differs from a generic Fermi-Dirac form, and is 
 computed with the public code {\it sterile-dm} of Venumadhav (2016; for an extended analysis see also Ghiglieri \& Laine 2015). The computation is based on the Boltzmann equation and includes detailed calculations of the lepton asymmetry around the quark-hadron transition. To obtain the power spectrum, the publicly available Boltzmann solver {\it CLASS} (Blas, Lesgourgues, \& Tram 2011) has  been used (Menci et al. 2017). 
\item RP2 model. As above, but with $\sin^2(2\theta)=5\,10^{-11}$, corresponding to $L_6\approx 10$. The spectrum is computed with the same tools described above.
\item SD1: Scalar Decay Model with $m_{\nu}=7$,  freeze-in regime (see Konig et al. 2016 for an extended discussion). Production from scalar decay is described by a generic model that invokes one real scalar singlet $S$ and one sterile neutrino $N$ beyond the Standard Model. 
The interaction between the scalar and the sterile neutrino is encoded in $\mathcal{L} \supset - \frac{y}{2} S\overline{N^c}N$ where $y$ is a Yukawa-type coupling determining the decay rate of the scalar and hence controlling how fast the scalar decays into sterile neutrinos. If the scalar develops a non-zero expectation value $\langle S\rangle$ this leads to a Majorana mass $m_{\nu}=y\,\langle S\rangle$. For scalar with $\langle S\rangle$ in the GeV-TeV range, couplings 
$y\approx 10^{-9}-10^{-5}$ are required to have $m_{\nu}\sim $ keV. 
The scalar singlet couples to the Higgs doublet $\Phi$ via a \emph{Higgs portal} $\mathcal{L} \supset 2\lambda \left(\Phi^\dagger \Phi \right) S^2$ where $\lambda$ is a dimensionless coupling which determines the production rate of the scalar. In the limit of small Higgs portal couplings (log $\lambda\ll -6$), the scalar itself is produced by freeze-in (at least when $y$ is within the range explored here, see  Heikinheimo, Tenkanen and Tuominen 2017 for 
more extended cases) and is always strongly suppressed compared to its would-be equilibrium abundance. As a representative case of this class of models, we consider the case $y=10^{-8.5}$, $\lambda=10^{-8}$, with $m_S=100$ GeV which - although yielding an appreciably suppressed power spectrum with respect to CDM - is marginally consistent with existing bounds from X-ray observations and from structure formation (Merle \& Totzauer 2015).  
The momentum distributions are derived in Konig et al. (2016) and the power spectrum is calculated with the Boltzmann solver {\it CLASS} as in Menci et al. (2017). Although
\item SD2: Scalar Decay Model with $m_{\nu}=7$,  freeze-out regime. The framework is the same outlined above, but in this case  $\lambda$ is large enough to equilibrate the scalars, so that they will be subject to the well-known dynamics of freeze-out. We shall consider the case $m_S=100$ GeV, 
$y=10^{-8.5}$, $\lambda=10^{-5}$ as a borderline case (with substantial suppression of the power spectrum with respect to CDM, and still  marginally consistent with existing bounds, see Merle \& Totzauer 2015) representative of this class of models. 
 \end{itemize}
 
\begin{center}
\vspace{0.1cm}
\hspace{-0.2cm}
\scalebox{0.51}[0.51]{\rotatebox{-0}{\includegraphics{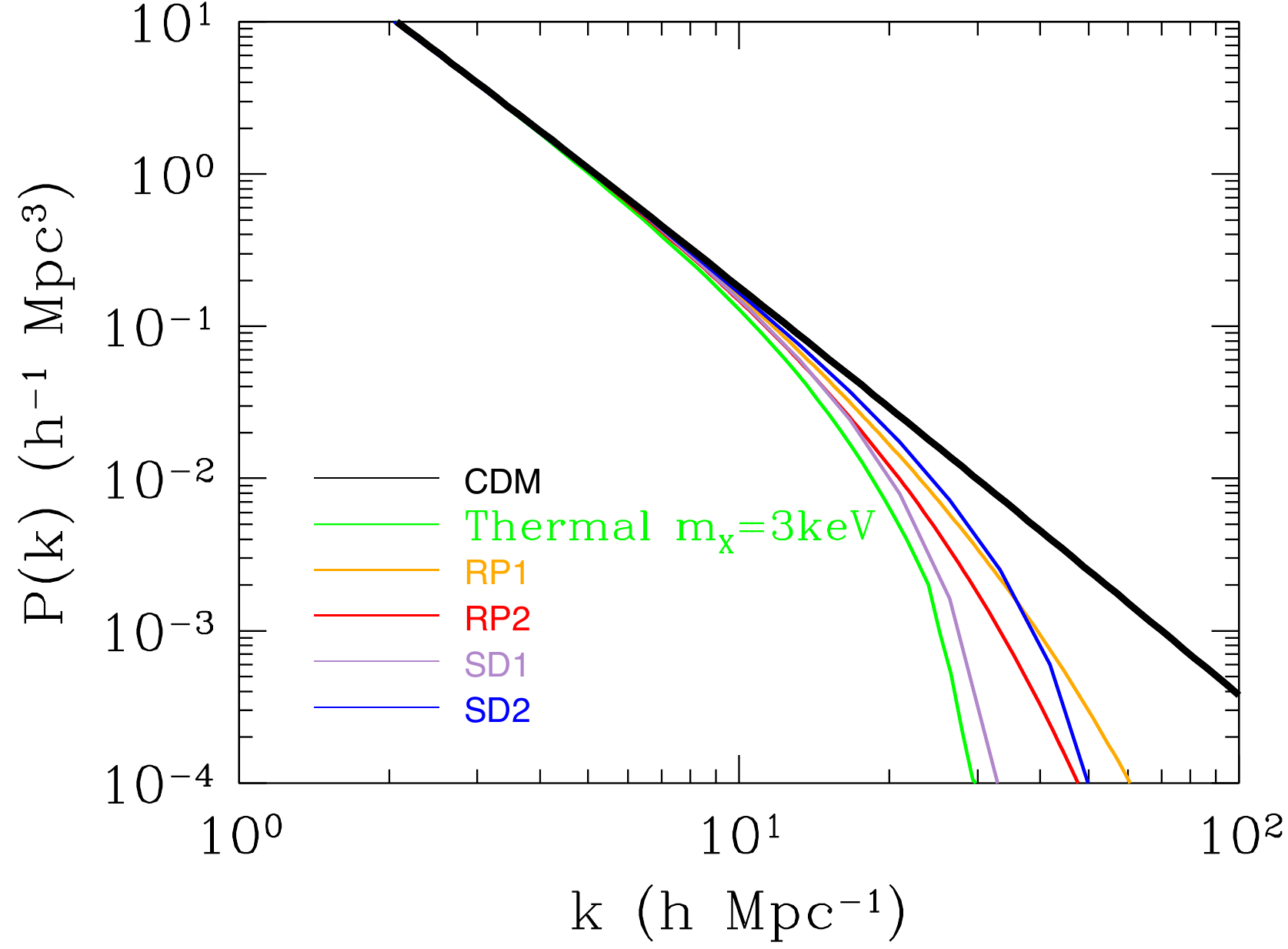}}}
\scalebox{0.5}[0.5]{\rotatebox{-0}{\includegraphics{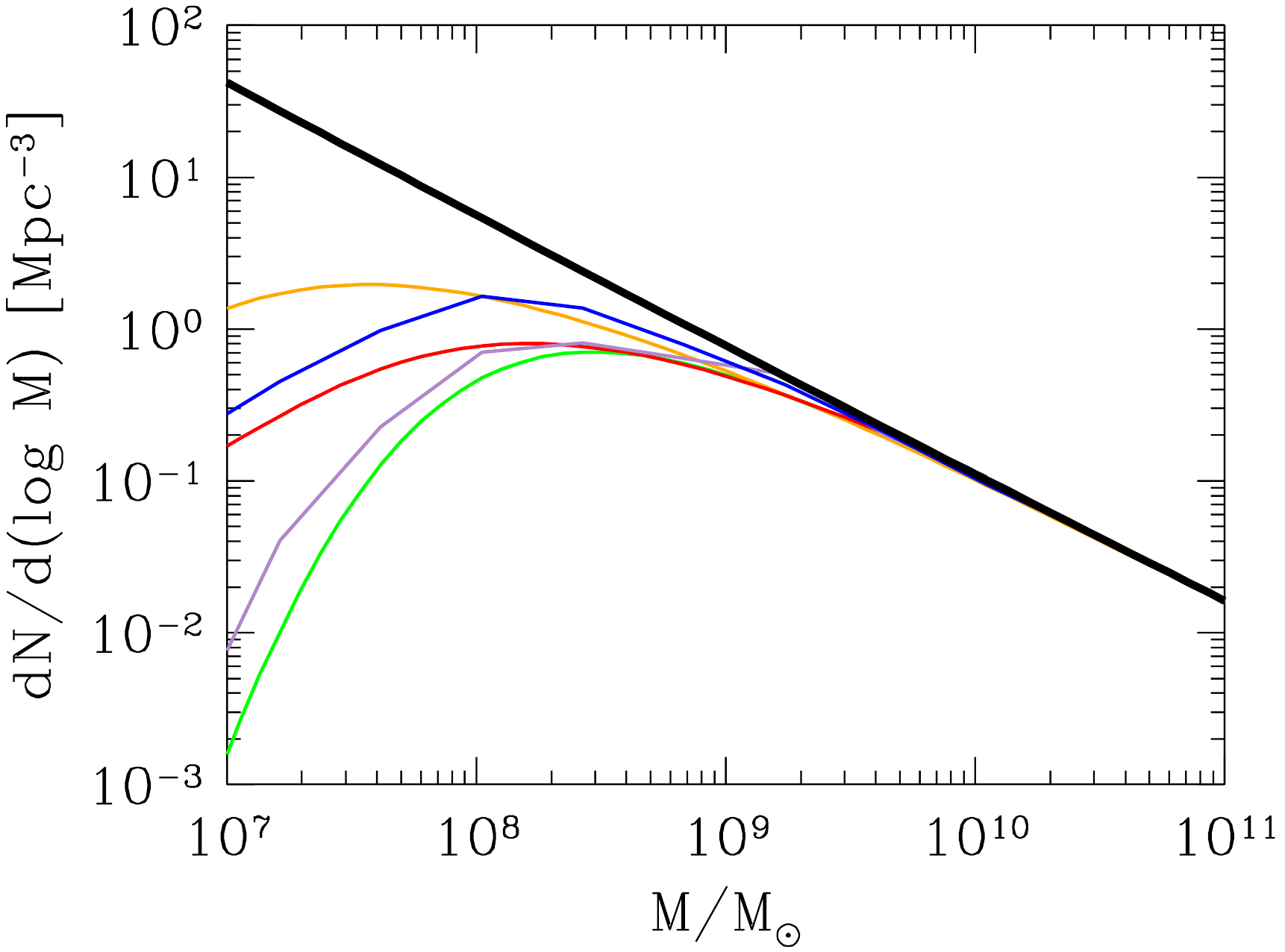}}}
\vspace{0.1cm }
\end{center}
{\footnotesize Figure 1. Left panel: The linear power spectrum at z=0 for the different DM models considered in the text, as indicated by the labes.
Right panel: for the same DM models, we show the DM halo mass function at z=0. }
\vspace{0.3cm }

The considered models are summarized in Table 1, where we  show key quantities characterizing the suppression of the power spectrum compared 
to the CDM case, i.e., the half-mode mass $M_{1/2}$, and the wavenumber $k_{peak}$ at which the dimensionless power spectrum  $k^{3}\,P(k)$ peaks. 
In the Table we also indicate the values of the parameters $v_0$ and $\alpha$ defining the Supernovae feedback adopted for each model, as described in 
detail in Sect. 2.3.
\newpage
\begin{center}
{\bf Table 1} \\
Considered Dark Matter models and corresponding key quantities
\end{center}

\resizebox{15.5cm}{!}{
%\begin{table}[]
\centering
%\caption{Dark Matter models}
\hspace{-1cm}

\begin{tabular}{lllcccc}
\hline
\textbf{Model} & \textbf{DM particle}                                                                                             & \textbf{Model parameters}                                                                                                                                                                                                       & \begin{tabular}[c]{@{}c@{}}$M_{hm}$\\ $(10^8\,h^{-1}\,M_{\odot})$\end{tabular} & \begin{tabular}[c]{@{}c@{}}$k_{peak}$\\ ($h\,$Mpc$^{-1}$)\end{tabular} & \begin{tabular}[c]{@{}c@{}}$v_0$\\ (km/s)\end{tabular} & $\alpha$ \\ \hline
\textbf{CDM}   & Cold relics                                                                                                      &                                                                                                                                                                                                                                 & \multicolumn{1}{l}{}                                                           & \multicolumn{1}{l}{}                                                   & 300                                                    & 3.3 \\ \hline
\textbf{WDM}   & Warm thermal relics                                                                                              & Relic particle mass $m_X$=3 keV                                                                                                                                                                                                 & 4                                                                              & 7.6                                                                    & 340                                                    & 2   \\ \hline
\textbf{RP1}   & \begin{tabular}[c]{@{}l@{}}Resonantly produced\\ Sterile Neutrinos. \\ Large mixing angle.\end{tabular}          & \begin{tabular}[c]{@{}l@{}}Sterile neutrino mass $m_{\nu}=7$ keV\\ Mixing parameter $\sin^2(2\theta)=2\,10^{-10}$\end{tabular}                                                                                                  & 2.6                                                                            & 10.7                                                                   & 360                                                    & 2   \\ \hline
\textbf{RP2}   & \begin{tabular}[c]{@{}l@{}}Resonantly produced\\ Sterile Neutrinos. \\ Small mixing angle.\end{tabular}          & \begin{tabular}[c]{@{}l@{}}Sterile neutrino mass $m_{\nu}=7$ keV\\ Mixing parameter $\sin^2(2\theta)=5\,10^{-11}$\end{tabular}                                                                                                  & 2.9                                                                            & 8.9                                                                    & 340                                                    & 2   \\ \hline
\textbf{SD1}   & \begin{tabular}[c]{@{}l@{}}Sterile Neutrinos produced\\ via scalar decay \\ in a freeze-in regime.\end{tabular}  & \begin{tabular}[c]{@{}l@{}}Sterile neutrino mass $m_{\nu}=7$ keV\\ Scalar coupling with Higgs sector $\lambda=10^{-8}$ \\ Scalar coupling with sterile neutrino $y=10^{-8.5}$\\ Scalar particle mass $m_S=100$ GeV\end{tabular} & 3                                                                              & 9.1                                                                    & 340                                                    & 2   \\ \hline
\textbf{SD2}   & \begin{tabular}[c]{@{}l@{}}Sterile Neutrinos produced\\ via scalar decay \\ in a freeze-out regime.\end{tabular} & \begin{tabular}[c]{@{}l@{}}Sterile neutrino mass $m_{\nu}=7$ keV\\ Scalar coupling with Higgs sector $\lambda=10^{-5}$\\ Scalar coupling with sterile neutrino $y=10^{-8.5}$\\ Scalar particle mass $m_S=100$ GeV\end{tabular}  & 1.6                                                                            & 13.5                                                                   & 360                                                    & 2   \\ \hline
\end{tabular}

}

\subsection{The semi-analytic model: the dark matter sector}

The backbone of the computation is constituted by the collapse history of DM halos on progressively larger scales. Realizations of such histories are generated through a Monte Carlo procedure on the basis of the merging rates given by the Extended Press \& Schechter (EPS) theory, see Bond et al. (1991); Bower (1991); Lacey \& Cole (1993). In this framework, the  evolution of the DM condensations is determined by 
the  power spectrum $P(k)$, that is computed for the considered DM models as described in sect. 2.1, through the variance $\sigma$ of the primordial DM density field. This is a function of the mass scale $M\propto \overline{\rho} r^3$ of the DM density perturbations (and of the background density $\overline{\rho}$) given by:
\begin{equation} 
\sigma^2(M)=\int {dk\,k^2\over 2\,\pi^2}\,P(k)\,W(kr)
\end{equation}
where $W(kr)$ is the window function (see Peebles 1993). While for CDM a top-hat shape in the real space is the canonical  choice for the filter function, both theoretical arguments (Benson et al. 2013; Schneider et al. 2013) and numerical experiments (see Schneider et al. 2012; 2013; Angulo et al. 2013) show that 
 the proper choice for models with suppressed power $P(k)$ at large $k$ is constituted by a sharp-$k$ form (a top-hat sphere in Fourier space). 
 While for a top-hat filter the mass assigned to the scale $r$ is $M = 4\pi\,\overline{\rho}\,r^3/3$, in the case of sharp-$k$ filter the mass 
assigned to the filter scale is calibrated with simulations. These show that adopting the relation $M = 4\pi\,\overline{\rho} (cr)^3/3$ with $c = 2.7$ (Schneider et al. 2013) the resulting mass distributions provide an excellent fit to N-body results for a wide range of DM masses and redshifts (see Schneider 2015). 

The differential halo mass function of DM halos (per unit $\log\,M$) can be calculated basing on the extended Press \& Schechter approach (Benson et al. 2013; Schneider 2013):
\begin{equation}
 {d \phi \over d  ln M}={1\over 6}\,{\overline{\rho}\over M}\,f(\nu)\,{d ln\,\sigma^2\over d ln  r}=-{1\over 2\,\pi^2\,\sigma^2(r)}\,{P(1/r)\over r^3},
\end{equation}
where the latter equality applies when the variance is computed adopting a  sharp-$k$ filter. 
Here $\nu\equiv \delta_c^2(t)/\sigma^2$ depends on the linearly extrapolated density for collapse in the spherical model, $\delta_c(t)=1.686/D(t)$, and $D(t)$ is the growth factor of DM perturbations. We conservatively assume a spherical collapse model, for which $f(\nu)=\sqrt{2\nu/\pi}\,\exp(-\nu/2)$. The effect of assuming different power spectra (corresponding to the different DM models introduced in Sect. 2.1) on the differential halo mass function is shown in the bottom panel of fig. 1. 

The merging trees of DM halos are generated through a Monte Carlo procedure (as described in Menci et al. 2005) from  the conditional mass function, which gives the abundance of haloes per mass $M$ and cosmic time $t$, eventually ending up in a single host halo with mass $M_0$ at final time $t_0$. In the case of sharp-$k$ filter (adopted for our models with suppressed power spectra) this reads (Benson et al. 2013; Schneider 2015) 
$$
{d N(M|M_0)\over d  ln M}={1\over 6\,\pi^{2}}\,{M_0\over M}\,f(\delta_{c},M|\delta_{c,0},M_0){P(1/r)\over r^3}
$$
where the conditional first-crossing distribution in the case of spherical collapse is given by 
$$
f(\delta_c,S|\delta_{c,0},S_0)={\delta_c(t)-\delta_{c,0} \over \sqrt{2\,\pi\,[\sigma^{2}(M) -\sigma^{2}(M_0)]} } \,e^{-{[(\delta_c(t)-\delta_{c,0}]^{2} \over 2\,\pi\,[\sigma^{2}(M) -\sigma^{2}(M_0)] } }. 
$$
The merging histories generated by the above Monte Carlo procedure allow to track the merging histories of  DM clumps down to the mass $M=10^7$ $M_{\odot}$, well below the half-mode mass scale characterizing the suppression in the power spectrum of the considered DM models with respect to CDM.

The dynamical evolution of sub-structures is computed in our Monte Carlo procedure as described in detail in Menci et al (2005, 2008). After each merging event, the dark matter haloes included into a larger object may survive as satellites, or sink to the centre due to dynamical friction to increase the mass of the central dominant galaxy. The  density profiles of DM haloes have been computed using a  Navarro, Frenk White (NFW 1997) form. The mass-dependence of the concentration parameter $c(M)$  has been taken from Macci\'o, Dutton \& van den Bosch (2008) for the CDM case. For the 
 DM models with suppressed power spectrum we computed $c(M)$ using the algorithm in Schneider et al. (2015), yielding a downturn of $c(M)$ for mass scales smaller than the half-mode mass scale $M\lesssim M_{1/2}$.

\subsection{The semi-analytic model: the baryonic sector and the setting of free parameters}

The baryonic processes taking place in each dark matter halo are computed as described in earlier works (see Menci et al. 2014, and references therein). The gas in the halo, initially set to have a density given by the universal baryon fraction and to be at the virial temperature, cools due to atomic processes and settles into a rotationally supported disk with mass $M_{gas}$, disk radius $r_d$, and disk circular velocity $v_d$, computed as in Mo, Mao \& White (1998). The cooled gas $M_{gas}$ is gradually converted into stars, with a star formation rate (SFR) $\dot M_*=M_{gas}/\tau_*$ given by the Schmidt-Kennicut law with a 
 gas conversion time scale $\tau_*=q\,\tau_d$ proportional to the  dynamical time scale $\tau_d$ through the free parameter $q$.  In addition to the above "quiescent" mode of star formation, galaxy interactions occurring in the same host dark matter halo may induce the sudden conversion of a fraction $f$ of cold gas into stars on a short time-scale $\sim 10^7-10^8$ yrs given by the duration of the interaction. The fraction $f$ is related to the mass ratio and to the relative velocity of the merging partners as described in Menci et al. (2003).  
The energy released by the supernovae associated to the total star formation returns a fraction of the disk gas into the hot phase,
at a rate $\dot M_h=\dot M_*/(v_d/v_0)^{\alpha}$ parametrized (as in most SAMs, see, e.g., Cole et al. 2000; Benson et al. 2003; Gonzalez-Perez 2014; see also the review  by Somerville \& Dav\'e 2015) in terms of the free parameters $v_0$ and 
$\alpha$, defining the normalization and the scaling of the feedback efficiency with the size of the host DM halo, respctively.  
Following existing works based on SAMs (see, e.g, Benson et al. 2003; Font et al. 2011) we model reionization feedback using a simple approximation in which dark matter haloes with circular velocity at the virial radius $v\leq  v_{crit}$  have no gas accretion or gas cooling at redshifts smaller than that corresponding to   reionization $z\leq z_{cut}$. We take $v_{crit}=25$ km/s (see, e.g., Okamoto, Gao \& Theuns 2008; see also Hou et al. 2016) and $z_{cut}=10$ (see, e.g. Benson et al. 2003,  Kennedy  2014); varying the assumed $z_{cut}$ in the range 7-10 does not appreciably change our main results). This simple model provides a good approximation to  more complex, self-consistent photoionization feedback models (Benson et al. 2002, Font et al. 2011), and is widely adopted in SAMs (see, e.g. Hou et al. 2016) including recent works on the comparison between CDM and WDM predictions (see Kennedy et al. 2014). 
 
An additional source of feedback is provided by the energy radiated by the Active Galactic Nuclei (AGNs) which correspond to the active accretion phase of the  supermassive black hole at the centre of DM haloes; the detailed description of our implementation of the AGN feedback is given in Menci et al. (2008); this, however, mainly affects the massive galaxy population, which does not constitute our main focus here. Finally, the luminosity - in different bands - produced by the stellar populations of the galaxies are computed by convolving the star formation histories of the galaxy progenitors with a synthetic spectral energy distribution, which we take from Bruzual \& Charlot (2003) assuming a Salpeter IMF. The model also includes a
treatment of the transfer of stellar mass to the bulge during mergers and the tidal stripping of stellar material from satellite galaxies, as described in detail in Menci et al. (2014).

The main free parameters in the modes are the normalization of the star formation timescale $q$, and the feedback normalization and scaling, $v_0$ and $\alpha$. For each considered DM model, we choose the first so as to  reproduce the observed correlation between the star formation rate and the gas mass  (fig. 2, left panels), while $v_0$ and $\alpha$ are calibrated so as to match the shape of the low-mass end of the local stellar mass function (fig. 2, right panels). 

\begin{center}
\vspace{-0.2cm}
\hspace{-0.2cm}
\scalebox{0.42}[0.42]{\rotatebox{-0}{\includegraphics{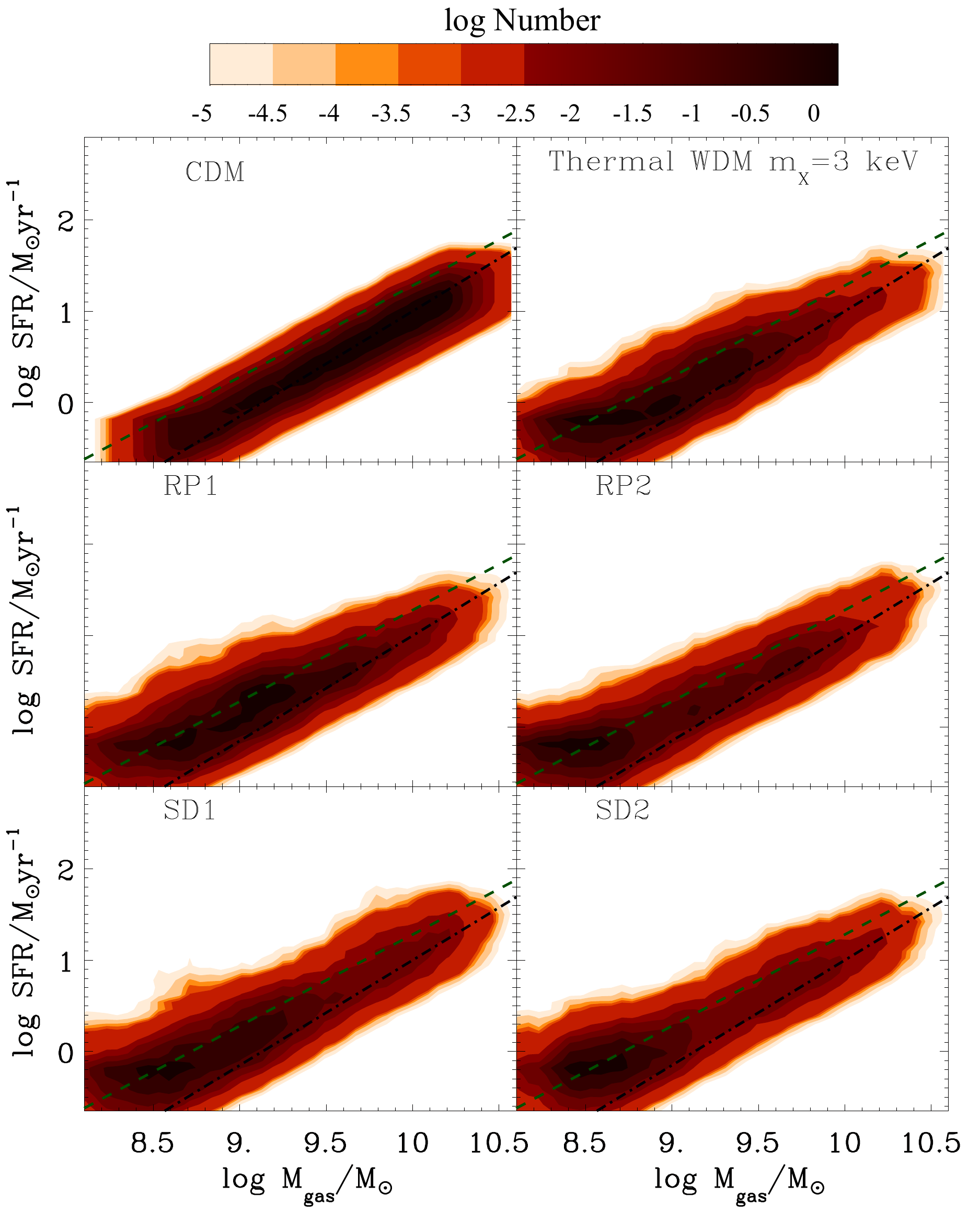}}}
\scalebox{0.42}[0.42]{\rotatebox{-0}{\includegraphics{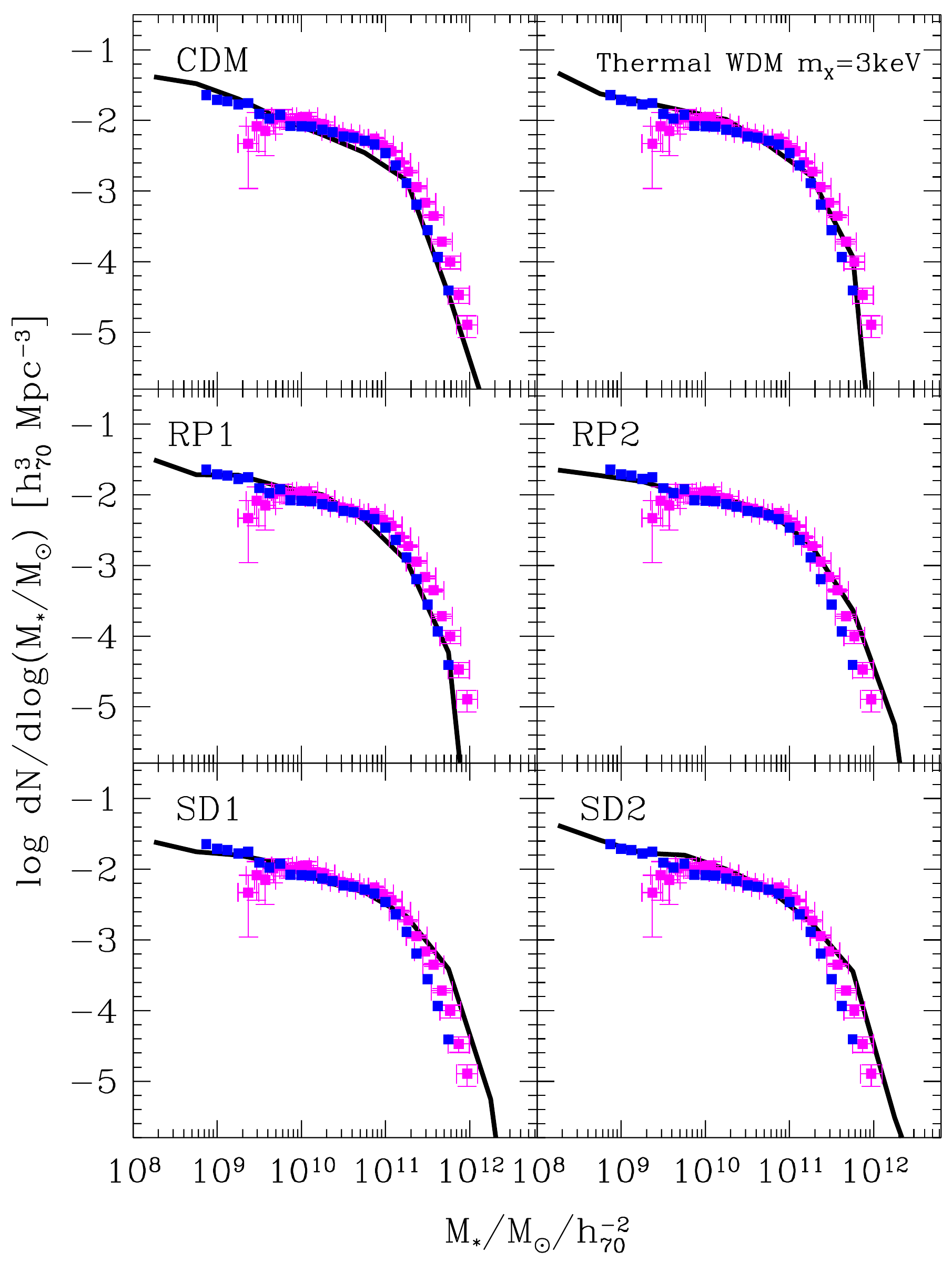}}}
\end{center}
{\footnotesize Figure 2. Left Panels: The relation between the star formation rate and the disk gas component: the contours show the distribution of 
model galaxies for the different DM models indicated by the labels. The lines correspond to the fit relation  given in eq. 8 of 
 Santini et al. (2014, dashed line) and in Genzel et al. (2010, dot dashed). The color code corresponds to the logarithm of the number of galaxies in each point of the 
 SFR-$M_{gas}$ plane normalized to the maximum value, as shown by the color bar.\newline
Right Panels: The local stellar mass function obtained from model galaxies in the different DM scenarios (solid line) are compared with data from Baldry et al. (2012, blue squares) and Li \& White (2009, pink squares).}
\vspace{0.1cm}

We find that we can keep the same star formation efficiency adopted in our earlier works $q=20$ for all  the different DM models, while matching the local stellar mass function (fig. 2, right panels) requires different values of the feedback parameter for the different DM models: while for CDM we take $\alpha=3.3$ and $v_0=300$ km/s, for the models with suppressed power spectrum we take $\alpha=2$ with $v_0$/km/s$=340, 360, 340, 340, 360$ for models  WDM, RP1, RP2, SD1, SD2, respectively. This is because the lower abundance of low-mass halos in WDM and SN models compared to CDM allows us to match the flat logarithmic slope of the faint end of the observed stellar mass function with a milder feedback (a combination which seems to provide a  slightly better match to the data compared to CDM, although the precise fit depends on the details of the baryon physics). The different combinations 
of feedback parameters adopted for the considered DM models are summarized in the rightmost columns of Table 1.

Our approach and the values of the feedback free parameters are  similar to those adopted in previous works (see, e.g., Kennedy et al. 2014), and corresponds to implementing weaker stellar feedback in 
DM models with suppressed power spectrum, since the lower abundance of low-mass galaxies in such models allows to match the local stellar mass 
function with a milder suppression of the $L/M$ (or $M_*/M$) ratio. Thus, matching the abundance of low-mass halos results into larger effective star formation 
in low-mass halos in models with suppressed power spectrum, as found in previous works based on abundance matching technique (see Corasaniti et al. 2017). In the following, we will investigate to what extent the different observables we compare with  enable us to disentangle the effects of 
feedback from those related to the assumed DM spectrum. 

\section{Results}
Basing on the SAM described above, we compute the observable properties of galaxies at low and high redshift and we compare with available data focusing our comparison on the low-mass end ($M\le 10^{9}\,M_{\odot}$) of the galaxy distribution, the one more  affected by the suppression in the power spectrum yield by the SN DM models. We aim at investigating to what extent the comparison with the different  observables can help to disentangle the effects of baryon physics (in particular of  feedback) from the specific effects of the different assumed DM models, and to single out the observational properties of low-mass galaxies that can potentially constitute a robust probe to strongly support or to rule out the considered SN models. 

\subsection{ Local properties of DM halos}

We first compare the predictions of the considered DM models with the abundance of satellites of Milky Way-like galaxies (fig. 3). In fact, this has long been constituting a major issue for CDM (Klypin et al. 1999; Moore et al. 1999), due to the large number of predicted sub-halos compared with the observed number of visible satellite galaxies. However, dwarf, satellite galaxies are also quite sensitive to the effect of baryonic feedback to  supernovae and to 
the effect of UV background. The potential of these processes to bring the abundance of Milky-Way like  CDM halos into agreement with observations has already been demonstrated using semi-analytical (Benson et al. 2002; Somerville 2001; Font et al. 2011; Guo et al. 2011) 
and hydrodynamical N-body models (see, e.g., Nickerson et al. 2011; Shen et al. 2014; Wheeler et al. 2015; Sawala et al. 2016).  
 The results of our SAM for the CDM case (see fig. 3, top left panel) confirm that the strong feedback adopted for the CDM model (Sect. 2.3) can bring the predicted abundance of low-mass luminous satellites close to the observed value. 
 Specifically,  we considered the predicted stellar mass function of 
 Milky Way-like halos, selected as those with DM mass in the range $M=1-2.5\cdot\,10^{12}\,M_{\odot}$ (where $M$ is the mass of a spherical volume with mean density equal to 200 times the critical density), which covers the bulk of the observational estimates of the Milky Way mass (for a comprehensive account of such measurements, see Wang et al. 2015). For the  CDM case we find that the predicted stellar mass function 
deviate by less than 1-$\sigma$ from the observed stellar mass function of M31, while still slightly overestimating the observed abundances of Milky Way satellites. The latter, however, should be considered as an effective lower limit due to the limited sky coverage of local galaxy surveys and the low surface brightness of dwarf galaxies (while the satellite counts of M31 should be closer to completeness for $M_*\geq 10^{5}\,M_{\odot}$).

\begin{center}
\vspace{-0.1cm}
\hspace{-0.2cm}
\scalebox{0.46}[0.46]{\rotatebox{-0}{\includegraphics{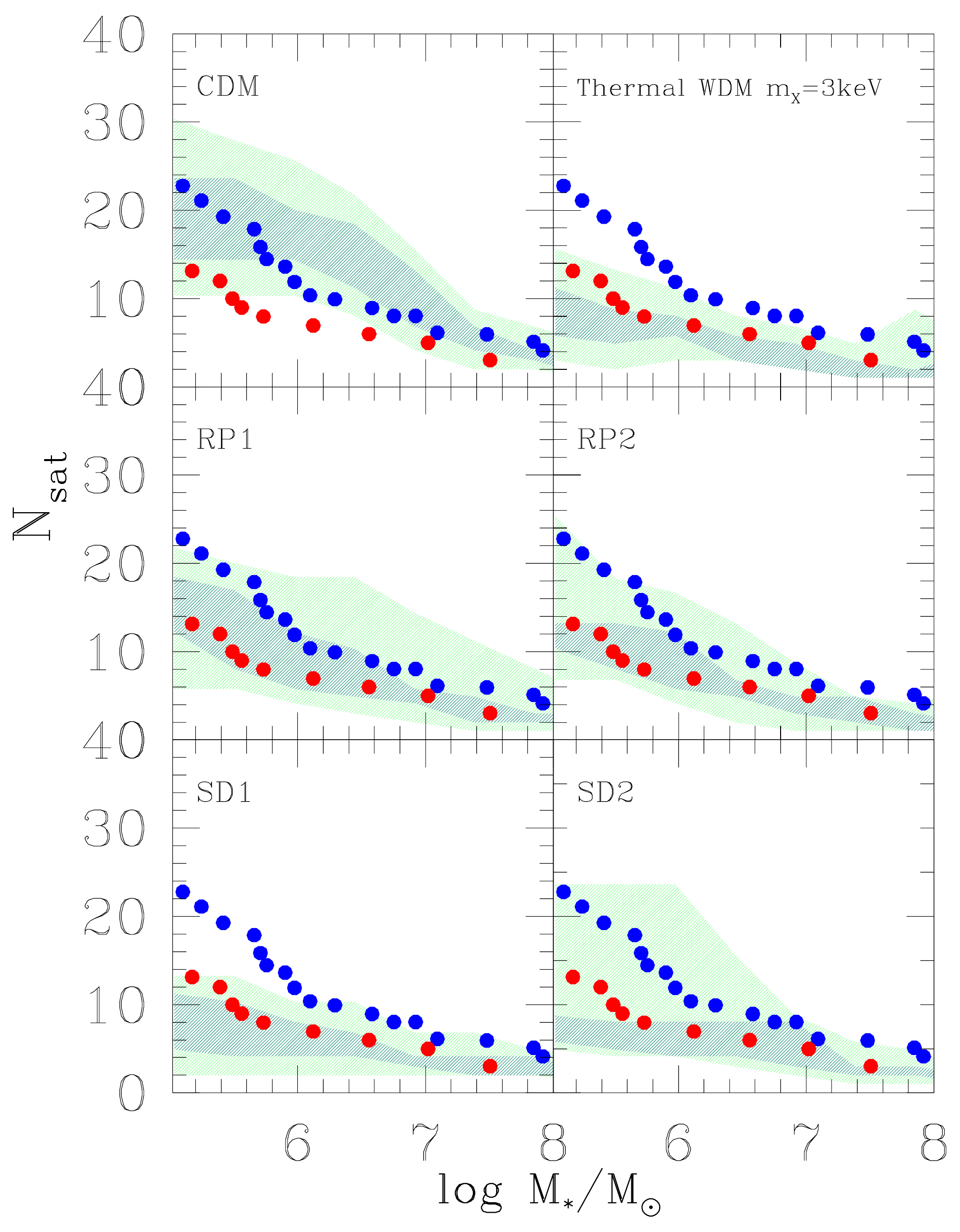}}}
\vspace{0.cm }
\end{center}
{\footnotesize Figure 3. The stellar mass function of satellites of Milky Way-like galaxies in the models are represented as shaded regions enclosing 
68 \% (darker) and 95 \% (lighter) of the satellites of halos with DM mass in the range $M=1-2.5\cdot\,10^{12}\,M_{\odot}$. These are compared with the compilation of 
observational data by McConnachie (2012) for the Milky Way satellites (red dots) and for the satellites of M31 (blue dots). }
\vspace{0.2cm }

The observed abundance of satellites provides the indication of a tension with the prediction of the thermal $m_X=3$ keV WDM model and of the SD1 model which underestimate the observed numbers. These indeed are the models characterized  by a stronger suppression of the power spectrum compared to CDM (see fig. 1); in the case of the thermal WDM model with $m_X=3$ keV, such a conclusion agrees with what found by Kennedy et al. (2014). 
As for the RP scenario, our RP1 and RP2 models are close to LA8 and LA12 models explored by Lovell et al. (2017a), corresponding to RP models with $m_{\nu}=7$ keV and 
lepton asymmetry $L_6=8$ and $L_6=12$, respectively. For such models we find consistent results, since both RP1 and RP2 match the observed satellite distributions, as obtained by Lovell et al. (2017a) for the LA8 and LA12 models. The convergence  of our results with existing  works on the impact of RP models on the Milky Way satellites is encouraging, and supports the robustness of 
our conclusions. Indeed, we have verified that such conclusions are not changed when more elaborated scaling laws for the SNae feedback (like the 
evolving feedback model in Hou et al. 2016) are assumed, or when the assumed value for $z_{cut}$ (Sect. 2.3) is varied in the range $z_{cut}$=7-10.

In principle, the degeneracy between feedback effects and the effects of assuming DM models with suppressed power spectra affecting the comparison with the  stellar mass function of satellite galaxies could be broken by investigating the stellar-to-halo mass ratios predicted by the different DM models.
In fact, we expect the strong feedback  needed in CDM to match the observed shape of the stellar mass distributions to yield lower $M_*/M$ ratios 
at small mass scales compared to the WDM and SN DM models. However, while $M_*/M$ ratios constitute a straightforward prediction of the models, 
on the observational side the measurements of the DM mass $M$ are subject to several uncertainties and biases. In fact, these are usually based 
on observed rotation velocities $v_{rot}$ measured through HI widths. However, the latter are related to the 
the maximum circular velocity of the dark matter halo $v_{max}$ (and hence to $M$) by relations depending on the assumed density profile which 
in turn depends on the assumed cosmology and on the feedback effects.  While high-resolution hydrodynamical simulations 
suggest a strong deviation of $v_{rot}$ from $v_{max}$ due  to strong stellar feedback (see, e.g., Macci\'o et al. 2016; Brooks et al. 2017) 
which provides shallower density distributions compared to the NFW form, 
 observationally-based estimates from HI rotation curves indicate a smaller difference (Read et al. 2017; Papastergis \& Shankar 2016; Trujillo-Gomez et al. 2016). Further uncertainties are introduced by the subtle procedure to obtain inclination-corrected HI profile half-width from the  observed line-of-sight rotation velocities (see, e.g., Papastergis et al. 2015).

Such a complex observational situation strongly affects the comparison with models, shown in fig. 4. To account for the uncertainties and biases affecting the 
observational determination of $M$, we compare the predictions of our DM models with the measurements from different groups that adopted different strategies. Ferrero et al. (2012) use HI rotation curves and stellar masses of galaxies compiled from the literature,  and 
use the outermost point of the rotation curve as a (conservative) proxy for $v_{max}$; in the cases of galaxies with peculiar rotation curves, they  use the velocity of the maximum of the rotation curve. Read et al. (2017) fit fully resolved HI rotation curves of individual field dwarfs  allowing for both a NFW profile and for parametrized cored profiles (accounting for baryon-induced cores), and apply the best fitting density profile to derive the DM halo mass $M$. 
 Brook and Di Cintio (2015) convert the observed stellar kinematics of 40 Local Group galaxies  to $M$ using the mass-dependent density profile derived from hydrodynamical simulations (Di Cintio et al. 2014), characterized by strong feedback effects; while not affected by issues related to the HI rotation curves, stellar kinematics only probe the very inner region of haloes, which are subject to large uncertainties in the halo mass estimates. 

We consider the above observational derivations of the $M_*/M$ ratio as representative of the different approaches used in the literature. E.g., 
Pace (2016) based on  Little THINGS and THINGS rotation curves fitted with the Di Cintio et al. (2014) halo profile. For larger halos  Katz (2017) use 147 rotation curves from the SPARC sample. Schneider, Trujillo-Gomez, Papastergis et al. (2017) follow a different approach to derive the DM
 mass. They select a sample of $109$ galaxies from the catalogue by Papastergis \& Shankar (2016) with rotation velocities measured to 
 radii  larger than three galactic half-light radius. This additional selection criterion guarantees that the velocity measurement is not dominated by baryonic effects. In addition they allow for cored profiles analogously to Read et al. (2017).  
 
We find that the strong Supernovae feedback required by the CDM model to match  the observed flat faint-end logarithmic slope of the  stellar mass distributions yields low $M_*/M$ ratio at the scale of dwarf galaxies: at mass scales $M\lesssim 10^{9}\,M_{\odot}$ we obtain $M_*/M\leq 10^{-5}$. This is a well-know feature of CDM models, as shown, e.g.,  by the blue line in fig. 4 representing the average ratios found by the full SAM model by Guo et al.  (2011) applied to large cosmological N-body simulations in the CDM case. A similar trend for CDM has been found using abundance matching techniques 
(Moster et al. 2010; Guo et al. 2010; Behroozi et al. 2013) or hydrodynamical simulations (Macci\'o et al. 2016).
Note that scatter in the predicted $(M_*/M)-M$ relation increases dramatically for decreasing masses $M\lesssim 10^9$ $M_{\odot}$.  Indeed, at the faintest end of the distribution, a  galaxy cannot be assigned a unique halo mass based solely on its stellar mass. Such a result is consistent with what  found in recent N-body simulations (see Munshi et al. 2017). 
 
\begin{center}
\vspace{-0.2cm}
\hspace{-0.2cm}
\scalebox{0.6}[0.6]{\rotatebox{-0}{\includegraphics{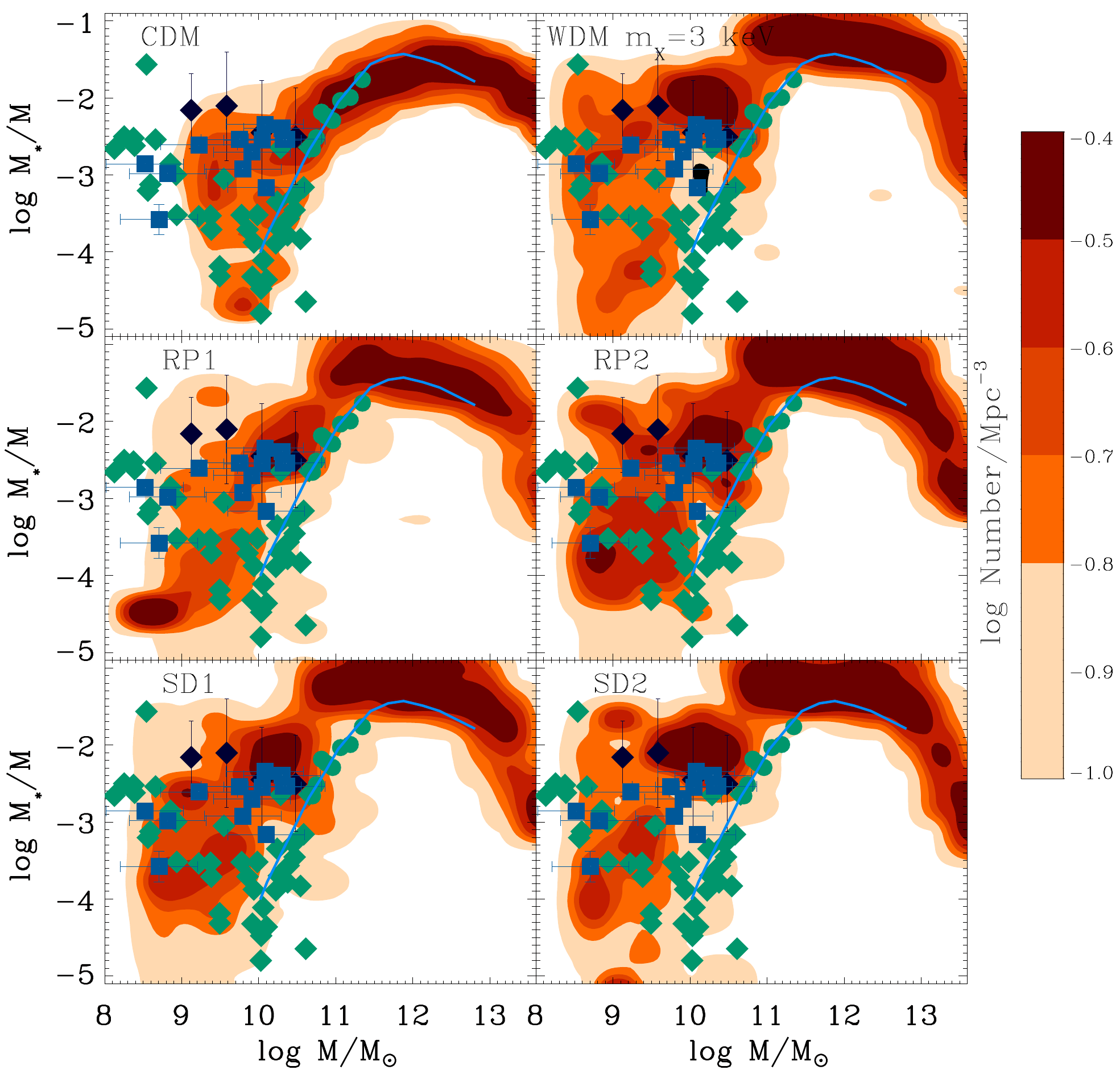}}}
\vspace{-0.2cm }
\end{center}
{\footnotesize Figure 4. The stellar-to-halo mass ratio $M_*/M$ as a function of the halo mass $M$ of model galaxies are  shown as coloured contours for the different assumed DM models. These are compared with different sets of data from Ferrero et al. (2012, black points), Brook and Di Cintio (2015, green points) and Read et al. (2017, blue squares). For the latter we excluded the Leo T data where no rotation curves are available. The blue line is the average relation obtained for CDM by  Guo et al. (2011) using a semi-analytic model applied to large cosmological N-body simulations.}
\vspace{0.2cm}

A different behavior characterizes the predictions from WDM and SN models. In these cases, $M_*/M$ ratios in the whole range $10^{-5}\lesssim M_*/M\lesssim 10^{-2}$ are found for low-mass halos $M\lesssim 10^9$ $M_{\odot}$. This is because the  suppression in the DM power spectrum at 
such mass scales allows us to match the observed stellar mass distributions with a milder feedback (see Sect. 2.1), thus yielding larger $M_*/M$ ratios, an effect  pointed out - for the thermal WDM case - in earlier works (see, e.g., Papastergis et al. 2015). While, in principle, the comparison of model predictions with the observed $(M_*/M)-M$ relation could constitute a sensitive probe for the DM scenarios, with the present data distribution better matched by WDM and SN models, the uncertainties outlined above results into  data that are too sparse to support any strong conclusions. Upcoming large area HI surveys with interferometric data, allowing for a detailed sampling of the rotational curves of low-mass galaxies will highly improve the observational situation in the next few years. 

\subsection{Evolution}

We now turn to investigate how assuming DM models based on SN affects the evolution of the galaxy population. We start by comparing the  evolution of the stellar mass function predicted by the considered SN models with the observed distributions (fig. 5). In this case we do not find any significant effect in the considered stellar mass range. Such  results are consistent with recent findings  obtained from SAMs coupled to high-resolution simulations for thermal WDM with $m_X=1.5$ keV (Wang et al. 2017). 
This is not unexpected, since the range of stellar masses $M_*\gtrsim 10^{8}\,M_{\odot}$ probed by the observations corresponds to DM halos $M\gtrsim 10^{9}\,M_{\odot}$ too large to be appreciably affected by the WDM and SN models considered here,  characterized by half-mode masses $M\lesssim 10^{9}\,M_{\odot}$ (although the observed range of stellar masses could probe WDM models with smaller values of $m_X\lesssim 1$ keV, corresponding to $M_{1/2}\approx 10^{10}\,M_{\odot}$, as shown in Menci et al. 2012). Pushing the comparison to smaller stellar masses $M_*\leq 10^{9}\,M_{\odot}$ constitutes a challenging task even in the near future, since at such masses (and especially at high redshifts) the derivation of the observed stellar mass is affected by corrections 
to account for the Eddington bias, which are not trivial to compute (see Grazian et al. 2015; Davidzon et al. 2017 for details). 
\begin{center}
\vspace{-0.1cm}
\hspace{-0.2cm}
\scalebox{0.65}[0.65]{\rotatebox{-0}{\includegraphics{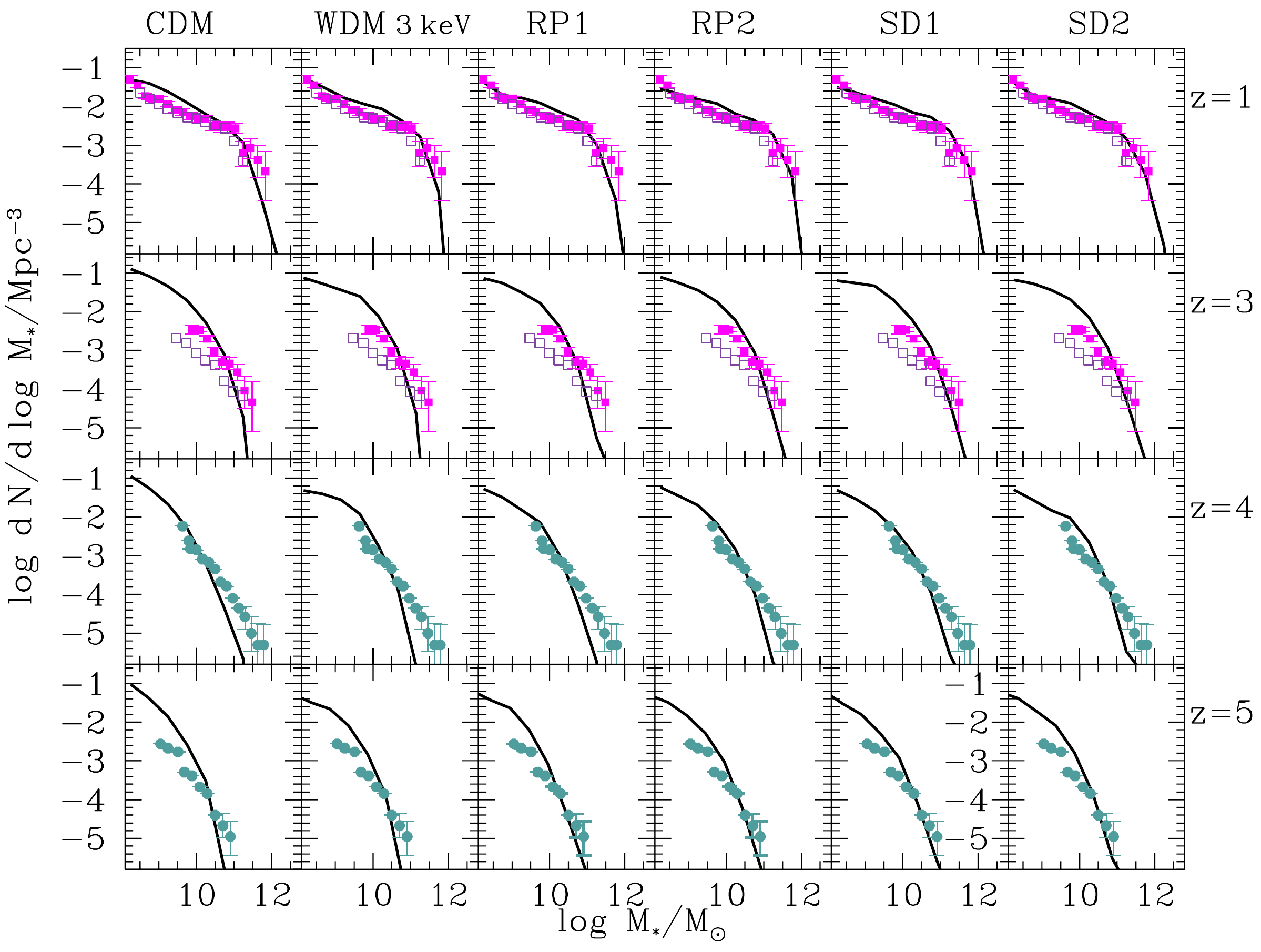}}}
\vspace{-0.5cm }
\end{center}
{\footnotesize Figure 5. The evolution of the predicted stellar mass function for the different DM models considered in the text and indicated at the top of the plot is shown for the different redshifts indicated on the right. Data for $z\leq 3$ are taken from Santini et al. (2012, filled squares) and Tomczak et al. (2014, open squares), while for $z\geq 4$ are taken from Grazian et al. (2015).}
\vspace{0.1cm}

A radically different situation holds when we compare with the observed UV luminosity functions up to very high redshifts $z=6$ (fig. 6). In fact,  the 
gravitational lensing magnification of background galaxies provided by foreground clusters  has been recently exploited in the framework of the Hubble Frontier Field (HFF) programme, leading to the to measurement the UV luminosity function down to extremely faint magnitudes $M_{UV}\approx -12$ at high redshifts $6\leq z\leq 8$ (Livermore, Finkelstein,  Lotz  2017).  Indeed, comparing the measured abundance of the faintest galaxies with   the maximum number density of DM halos in WDM cosmologies has already allowed to set robust limit $m_X\geq 2.4$ keV  independently of baryon physics for the mass of thermal relic WDM particles at 2-$\sigma$ confidence level (Menci et al. 2016), and to effectively constraint the parameter space of SN models (Menci et al. 2017). The above limits are very conservative, since they are derived comparing the observed number density of galaxies with the {\it maximum} abundance of  halos in different DM models; we thus expect that tighter constraints can be obtained when the observed abundances are compared with the  number density of luminous ($M_{UV}\lesssim -12$) galaxies predicted by different DM models. Such a comparison is performed in detail in fig. 6, where the UV luminosity functions of galaxies are compared with existing observations up to $z=6$. 

For the CDM case, we recover the long-standing problem of the over-production of low-luminosity galaxies at redshift $2-4$ (see, e.g., Somerville et al. 2001; Croton et al. 2006; Lo Faro et al. 2009;  Gruppioni et al. 2015), an instance of the CDM small-scale issues. 
On the other hand,  the lower abundance of low-mass galaxies characterizing models with suppressed power spectra provide a better fit to the UV luminosity functions up to 
redshift $z\approx 4$ (as already noted in Menci et al. 2012 for the thermal WDM case), but 
compares critically with the observed abundance at $z\approx 6$. This is indeed the most effective probe for the abundance of early forming, low-mass galaxies and - hence -  for the effects of suppressed power spectra. However, 
 although the number density derived by Livermore et al. (2017) is  robust from the statistical point of view, there could be subtle systematic effects related to the estimation of the survey volume, i.e., the variance of the lensing magnification maps of HFF clusters and the physical sizes of faint, high-z galaxies 
 (which enters the completeness correction). These have been claimed to affect the number density of high redshift galaxies in 
 the faintest bins, potentially leading to a flatter slope of the UV luminosity function at the faint end (Bouwens et al. 2017a,b; see also Kawamata et al. 2017). Thus, the conservative estimation of the UV luminosity function of Bouwens et al. (2017a) has also been shown in fig. 
 6, to provide an overview of the present observational situation. 

\begin{center}
\vspace{0.1cm}
\hspace{-0.2cm}
\scalebox{0.65}[0.65]{\rotatebox{-0}{\includegraphics{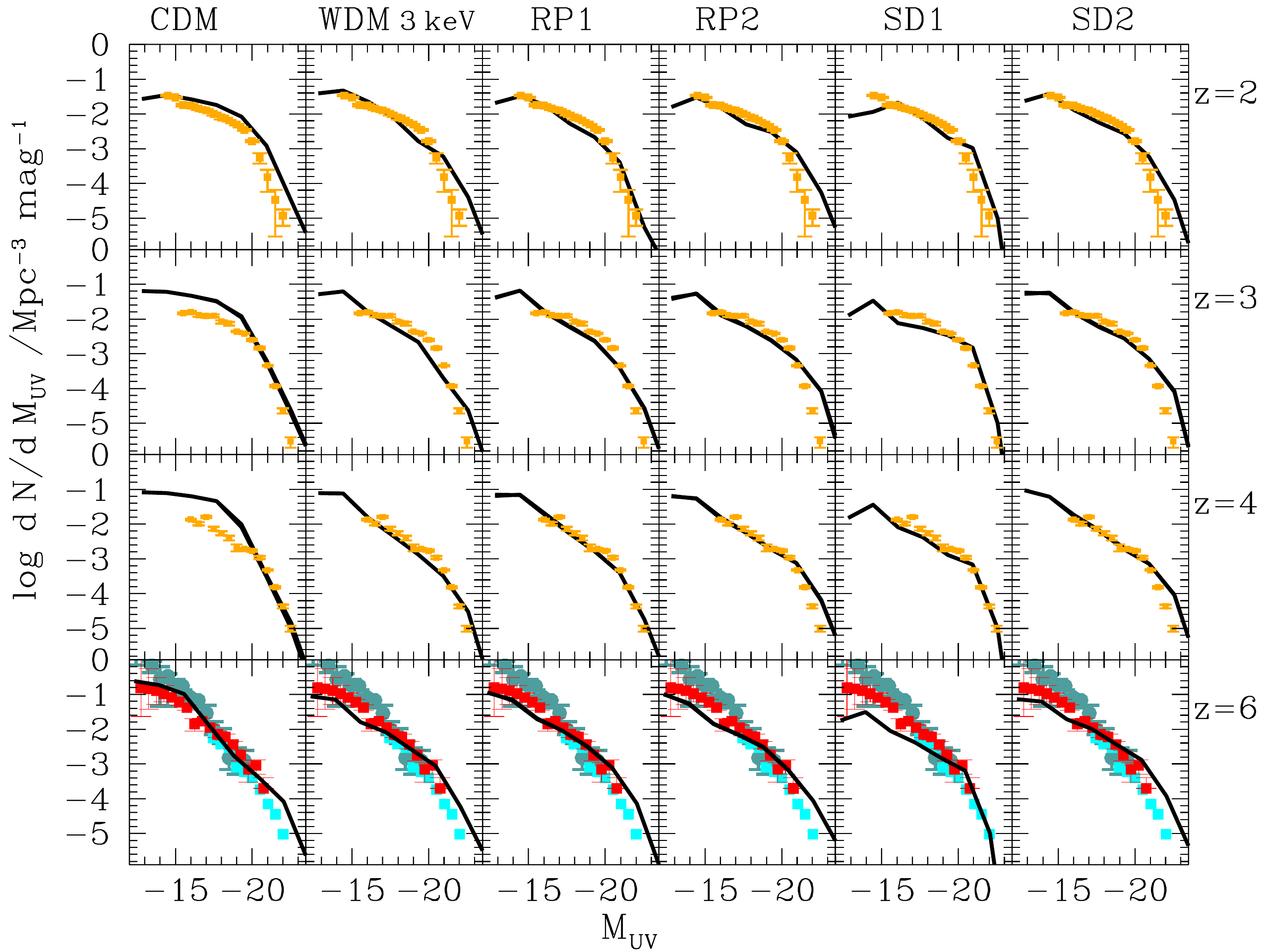}}}
\vspace{-0.2cm }
\end{center}
{\footnotesize Figure 6. The evolution of the predicted UV luminosity function for the different DM models considered in the text and indicated at the top of the plot is shown for the different redshifts indicated on the right. Data for $z\leq 4$ are taken from Parsa et al. (2016). Data in the highest redshift bin are taken from Livermore, Finkelstein \& Lotz (2017, dark green circles), from Bouwens et al. (2017a; red squares) for the faint end, while for the bright end we compare with Finkelstein et al. (2015a, light cyan squares). }
\vspace{0.2cm}

We obtain  that  all models with suppressed spectra underestimate the abundance of high-$z$ galaxies with $M_{UV}\gtrsim -14$ measured by Livermore et al. (2017); 
the two DM models with the most suppressed spectra (the thermal WDM case with $m_X=3$ keV and the SD1 model) also underestimate the 
more  conservative observational estimates (Bouwens et al. 2017a) of the abundance of $z=6$ faint galaxies, with a discrepancy $\gtrsim 1\sigma$. When models are compared with the Livermore et al. (2017) measurements, the WDM case with $m_X=3$ keV and the SD1 model underestimates the observed abundances by more than 2-$\sigma$, while the RP1, RP2, and SD2 models deviate by $\approx 1-\sigma$.
This shows that at present the UV luminosity functions at high-$z$ constitute an extremely  powerful probe for the DM scenarios. Thus, on the observational side, the first step to improve the results presented in this paper consists in a deeper understanding of the systematics associated with the lensing observations of faint, high-redshift galaxies. In fact, the analysis of the HFF observations is open to several advancements (see, e.g., Castellano et al. 2016). In addition, the present measurements of the UV luminosity functions from the HFF is based only on the first two fields of the HFF survey: the inclusion of the remaining four strong lensing clusters (Lotz et al. 2017) will reduce both statistical uncertainties and mitigate possible cosmic variance effects. In a few years from now a significant leap will be made possible by the availability of deep JWST imaging. In particular, the capability of reaching 30.5~AB (at $S/N=5$) in deep NIRCam fields (e.g., Finkelstein et al. 2015a,b)  will improve by~1.5 mags the depth of current HFF imaging, reaching absolute magnitudes of $M_{\rm UV}\approx -11$, and will yield 5 times larger samples of high-redshift galaxies (Laporte et al. 2015)  while significantly improving photometric selections through the availability of rest-frame optical photometry of high-$z$ sources.

\subsection{Star Formation}

Finally, we investigate the star formation properties of the galaxy populations in the different DM models considered here. In fig. 7 we plot the global star formation rate densities from our models and compare them with the observed values. The observational values  have been taken from the review paper by Madau \& Dikinson (2014) who collected a large set of  measurements in the literature, obtained either from UV and from FIR rest-frame observations (see caption). To perform a proper comparison, the 
model star formation densities have been obtained including all galaxies brighter that $0.03\,L_*$, the same threshold adopted for the observational values. All the models yield star formation rate densities consistent with 
observations. This is mainly because in the star formation density $\psi=\int N(\dot M_*)\,\dot  M_*\,d\dot M_*$ the number density $N(\dot M_*)$ of galaxies  is weighted with the star formation rate $\dot M_*$. While the first is suppressed for low-mass galaxies in all DM models with suppressed power spectrum, such an  effect is balanced by the larger $\dot M_*$ associated to these galaxies in such models due to the smaller adopted feedback efficiency (see sect. 2). Such a result shows that the star formation density does not constitute an effective probe for the different  DM scenarios, since in this case the DM and the baryonic effects are highly degenerate, at least in the luminosity range ($L*\geq 0.03\,L_*$) explored by present measurements. 
 
\begin{center}
\vspace{-0.1cm}
\hspace{-0.2cm}
\scalebox{0.55}[0.55]{\rotatebox{-0}{\includegraphics{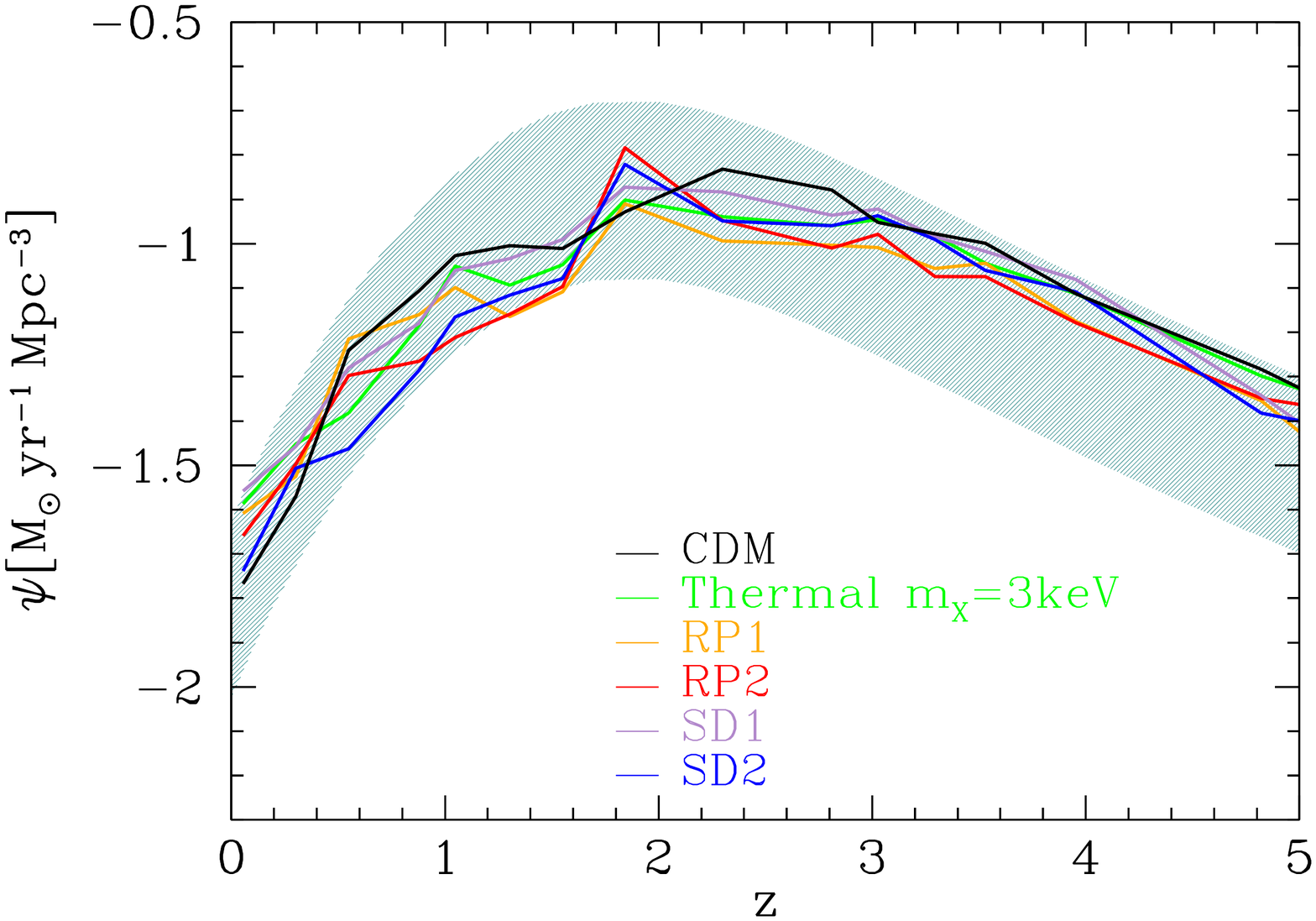}}}
\vspace{-0.2cm }
\end{center}
{\footnotesize Figure 7. We show the predicted star formation rate density for the different DM models considered in the text and shown by the labels.
The shaded region corresponds to the present uncertainties as result from the compilation of data in Madau \& Dickinson (2014), who consider 
only surveys that have measured SFRs from rest-frame FUV (generally 1500A°), MIR, or FIR measurements, and including all galaxies with luminosities 
larger than 0.03 $L_*$, where $L_*$ is the characteristic luminosity of the considered data sample. For the model predictions, we adopted the same lower luminosity cut, deriving $L*$ as the ratio between the second and the first moment of the UV luminosity functions. }
\vspace{0.2cm }
  
However, a deeper insight can be gained by investigating the star formation histories of low-mass galaxies in the different scenarios. In fig. 8 we show the stellar mass growth histories of low-mass galaxies for all the considered DM models. This is defined as the 
 stellar mass formed in all progenitors of a given galaxy by a given cosmic time $t$ normalized to the stellar mass of the galaxy at $z=0$. In the figure, we consider only low-mass galaxies (stellar masses $9\leq log M_*/M_{\odot}\leq 9.5$ at $z=0$), the ones affected by the assumed DM power spectrum. The large suppression in the number of progenitors of low-mass galaxies occurring in models with suppressed power spectrum overwhelms the larger star formation occurring in each progenitor due to milder feedback, and leads to an overall delay in the growth of the  stellar mass component of dwarf galaxies. The effective delay depends on the assumed DM scenario. Taking as a reference 
 value the time at which 80\% of the final stellar mass is formed in the progenitors (shown by the dashed lines in the figure), we obtain delays ranging from $\sim 1$ Gyr (in the case of thermal WDM) to $\sim 500$ Myr. Our findings are consistent with the existing results in the literature obtained for the WDM case. E.g., in the hydrodynamical simulations by Governato et al. (2015) a delay $\sim 0.5-1$ Gyr is found when comparing the stellar growth of a thermal WDM model 
 with $m_X=2$ keV with the CDM predictions.

\vspace{0.2cm}
\begin{center}
\scalebox{0.64}[0.64]{\rotatebox{-0}{\includegraphics{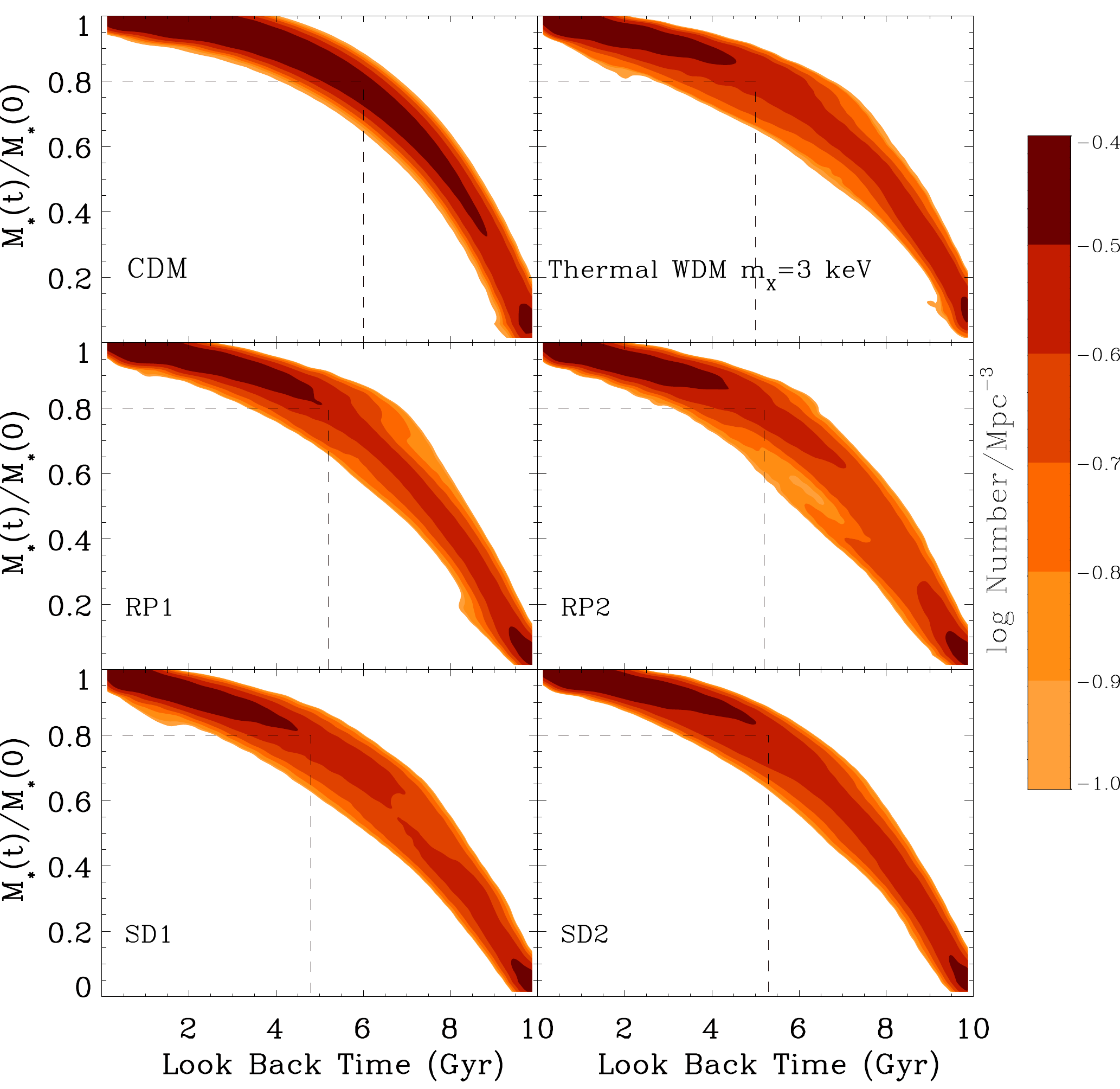}}}
\end{center}
{\footnotesize Figure 8.  The predicted stellar growth histories for the DM models indicated by the labels. The contours show, at given look-back time, the number of model galaxies with a given ration $M_*(t)/M_*(0)$ between  the mass formed in all progenitors of the final galaxies and the final stellar mass. Only galaxies with final mass $9\leq \log M_*/M_{\odot}\leq 9.5$ are considered. The dashed line show the look-back time corresponding to the formation of 80\% of the final stellar mass. }
\vspace{0.2cm}

Such an effect is particularly interesting, since it allows for observational tests of the DM scenarios using different indicators of the building-up of the stellar component in dwarf galaxies. An instance is provided by the specific star formation rate (SSFR) defined as $SSFR\equiv\dot M_*/M_*$. It is a measure of the present star formation activity normalized to the total amount of stars formed during the past history of star formation and mass assembly. In particular, values much smaller than the inverse of the Hubble time (typically $0.3/ t_H\approx 0.2\,10^{-10}$, see, e.g., Damen et al. 2009) correspond to quiescent galaxies, which must have formed most of their stellar mass at earlier times. Observational estimates of the full SSFR distributions based on the large statistics provided by the SDSS catalogue have been obtained by several authors (see, e.g.,  Balogh et al. 2004; McGee et al. 2011; Peng et al. 2012; Wetzel et al. 2012). Here (fig. 9) we compare with the distributions obtained by Wetzel et al. (2013) based on the spectroscopic NYU Value-Added Galaxy Catalog (NYU-VAGC, Blanton et al. 2005) from SDSS data release 7 (Abazajian et al. 2009) who constructed stellar mass limited samples complete down to stellar masses $M_*=5\,10^9$ $M_{\odot}$ and magnitudes $M_r=-19$. The measured SSFR are based on the spectral reductions by Brichmann et al. (2004) with updated prescriptions for AGN contamination, and are derived from emission lines for ${\rm SSFR}\geq 10^{-11}$ yr$^{-1}$, and from a combination of emission lines and $D_n4000$ for lower values of the SSFR. In the above sample of galaxies, Wetzel et al. (2013) have identified the objects that occupy the same host halo through a modified implementation of the group-finding algorithm of Yang et al. (2005; 2007); this allowed to obtain separate SSFR distribution for central and satellite galaxies (see Wetzel 2013 for details). 

The above observational distributions are compared with our predictions for the different DM models in fig. 9. 
To comply with the procedure adopted for observational distributions, model galaxies with log SSFR$/yr^{-1}\leq -12$ have been assigned a random Gaussian value centered at log SSFR$/yr^{-1}=-0.3\,log(M*/M_{\odot})-8.6$  and dispersion 0.5.
A clear bimodal distribution is found, in agreement with previous results from semi-analytic models based on simulations (Guo et al. 2011; Henriques et al. 2015). 

\begin{center}
\vspace{-0.3cm}
\scalebox{0.67}[0.67]{\rotatebox{-0}{\includegraphics{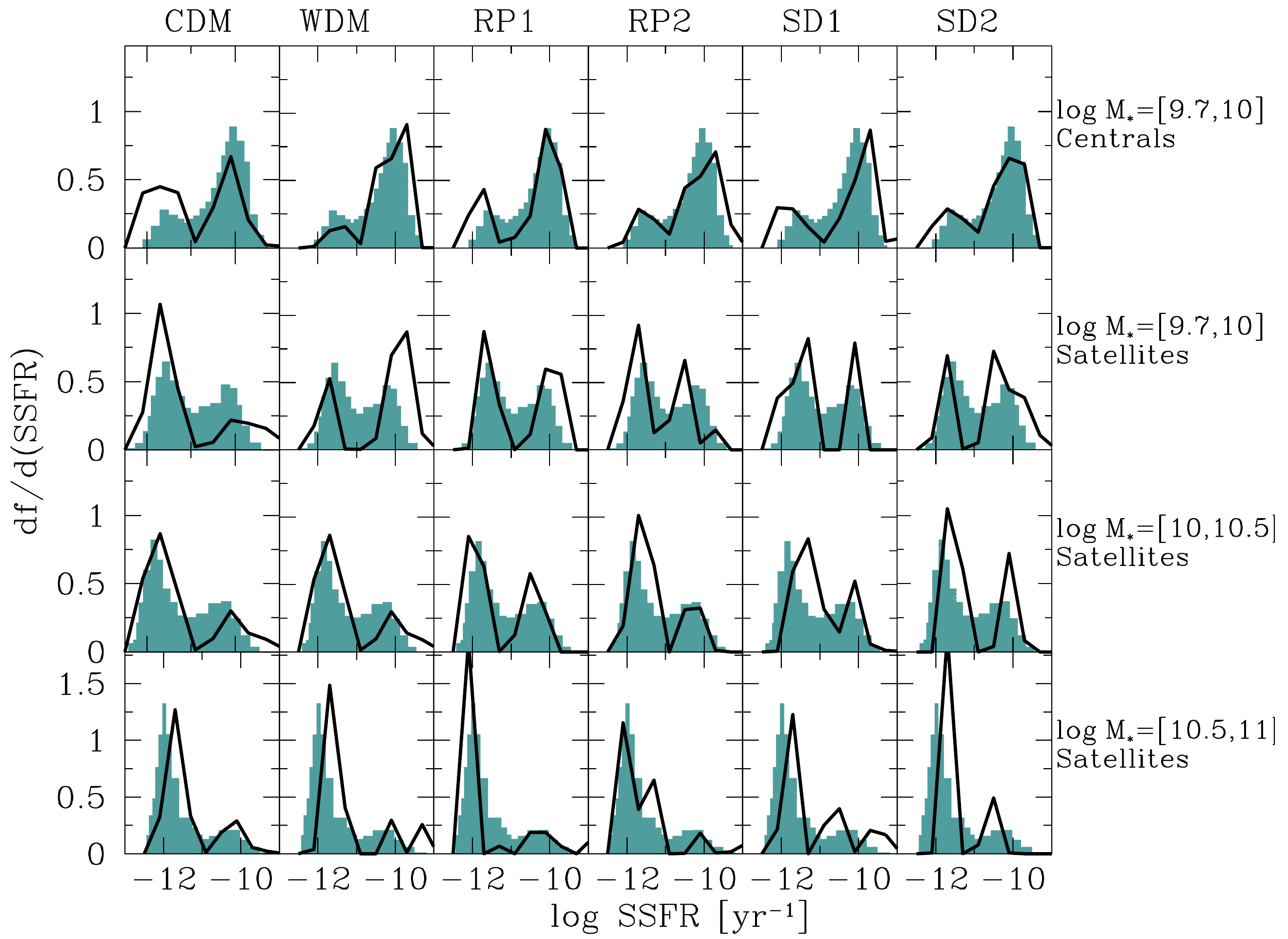}}}
\vspace{-0.4cm }
\end{center}
{\footnotesize Figure 9. Predicted distributions of specific star formation rate (solid lines) for central and satellite galaxies in different mass bins as 
indicated by the labels, for the different DM models indicated on the top.  The solid histograms are the observed distributions measured by Wetzel et al. (2013).
}
\vspace{0.2cm}

The detailed balance between passive and active galaxies and between environmental and internal quenching depends on the details of the implemented baryonic processes, including time-dependent strangulation effects (e.g., Guo et al. 2011) which we do not consider in this work. However, we find a systematic increase in the star-forming fraction of low-mass galaxies when models with suppressed power spectra are considered (in the first two rows strongest peaks at log SSFR$/yr^{-1}\approx -10$ are present in columns 2-6 when compared to the first column). 
 Such an effect  cannot originate only from environmental quenching processes affecting satellites since, to some extent, it is present also in central galaxies. Thus, it  
 must originate from the delays in the star formation histories of models with suppressed power spectra compared to CDM shown in fig. 9.  In such models, the stellar mass growth histories are skewed toward later cosmic times, yielding a larger fraction of active galaxies (with SSFR $\geq 10^{-11}$ yr$^{-1}$) compared to the CDM case. 
 
An useful diagnostic to probe the above effects of the adopted DM power spectrum on the star formation histories of galaxies is the cumulative 
age distribution of stellar populations in local galaxies. 
The importance of this topic was already shown in the work of Calura, Menci \& Gallazzi (2008), 
where they were comparing 
the age distribution of galaxies calculated in a SAM assuming a CDM cosmology to those computed within a 
WDM model, assuming that WDM was constituted by thermal relic particles of mass 0.75 keV.  \\
In Fig. 10 we show the observed cumulative age distribution of SDSS galaxies, which were first presented in  
Gallazzi et al. (2008), compared to the theoretical distributions computed assuming different DM scenarios. 
As explained in Calura et al. (2014), for the purpose of a fairer comparison between observations and models  
the mass-weighted ages of the stellar populations in local galaxies were re-calculated 
weighting each galaxy spectrum by 1/V$_{max}$, 
where V$_{max}$ is defined as 
the maximum visibility volume given by the bright and
faint magnitude limits of the observational dataset, $14.5\le r \le17.77$.\\
The steeper rise of the CDM age distribution reflects the extreme paucity of young galaxies characterising 
this model. 
All the models computed in the alternative cosmologies show a similar behaviour and a better agreement with the 
observed distribution than the CDM model. In general,  even if  the percentage of young galaxies is underestimated by all models, 
the striking feature is that all the models with suppressed power spectra show the presence of galaxies younger than $10^{9.3}$ yr, 
absent in the CDM model. 
Such a result is qualitatively in agreement with that 
obtained by Calura et al. (2014), although in such a work the assumed WDM power spectra was much more extreme and 
chosen in order to underline  the differences with the CDM scenario. 
In the future, it will be interesting to study how the ages of the stellar populations 
of local galaxies are distributed as a function of the stellar mass in all the different cosmological models considered in this 
work: this very aspect will be addressed in a forthcoming work. \\

\vspace{-0.4cm}
\begin{center}
\scalebox{0.6}[0.6]{\rotatebox{-0}{\includegraphics{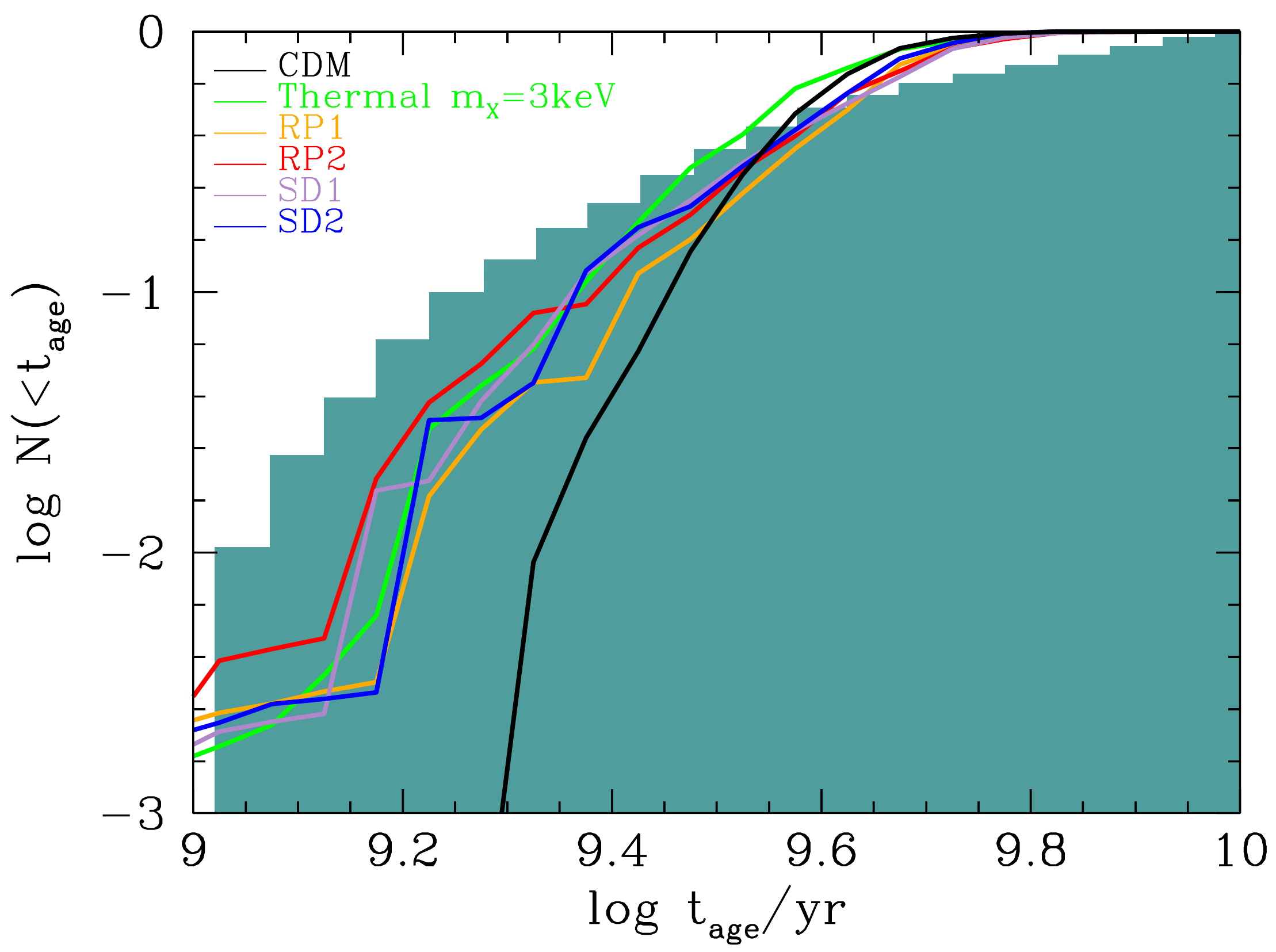}}}
\end{center}
\vspace{-0.2cm}
{\footnotesize Figure 10.  For the different DM scenarios considered in the text, we plot the predicted cumulative distributions of the 
mass-weighted age $t_{age}$ for the different DM models (solid lines), and compare them with the data by Gallazzi et al. (2008, solid histogram) weighted as described in Calura, Menci, Gallazzi (2014; see text). 
In the plot we have considered all galaxies with $M\geq 10^9\,M_{\odot}$.}
\vspace{0.2cm}

The enhanced presence of young, low-mass galaxies in the DM models with suppressed power spectra compared to the CDM case 
can be further investigated studying the integrated photometric properties of galaxies. \\
In the past, the integrated colours of composite stellar populations turned out as extremely useful to gain crucial clues on their formation history, as
shown in previous works addressing the star formation history of dwarf spheroidal galaxies (dSph) and dwarf irregulars
in the Local Group (Mateo 1998), a possible evolutionary connection between these two classes (e.g. Skillman \& Bender 1995), 
as well as the link of dSphs with large spheroids (Calura et al. 2008). \\
In Fig. 11 we show the distribution of (U-B) colors, 
plotted as a function of the absolute B-band magnitude, as computed by means of our SAM assuming different 
cosmologies and compared to an observational sample drawn from the SDSS dataset (Data Release 14). 
Here we focus mostly on the faintest systems, since, as already discussed, 
the major differences between CDM and WDM/SN spectra concerns mass scales $M\lesssim 10^{9}\,M_{\odot}$, i. e. 
halos which presumably host faint, dwarf galaxies. Thus, we  have extracted
from the entire SDSS sample of local ($z<0.1$) galaxies only those with with  stellar masses  $M_*\leq 10^{9.5}\,M_{\odot}$ \footnote{For the SDSS galaxies, the stellar masses have been calculated using the 
the Bruzual and Charlot (2003) stellar population synthesis models.}.

 \vspace{0.2cm}
\begin{center}
\scalebox{0.68}[0.68]{\rotatebox{-0}{\includegraphics{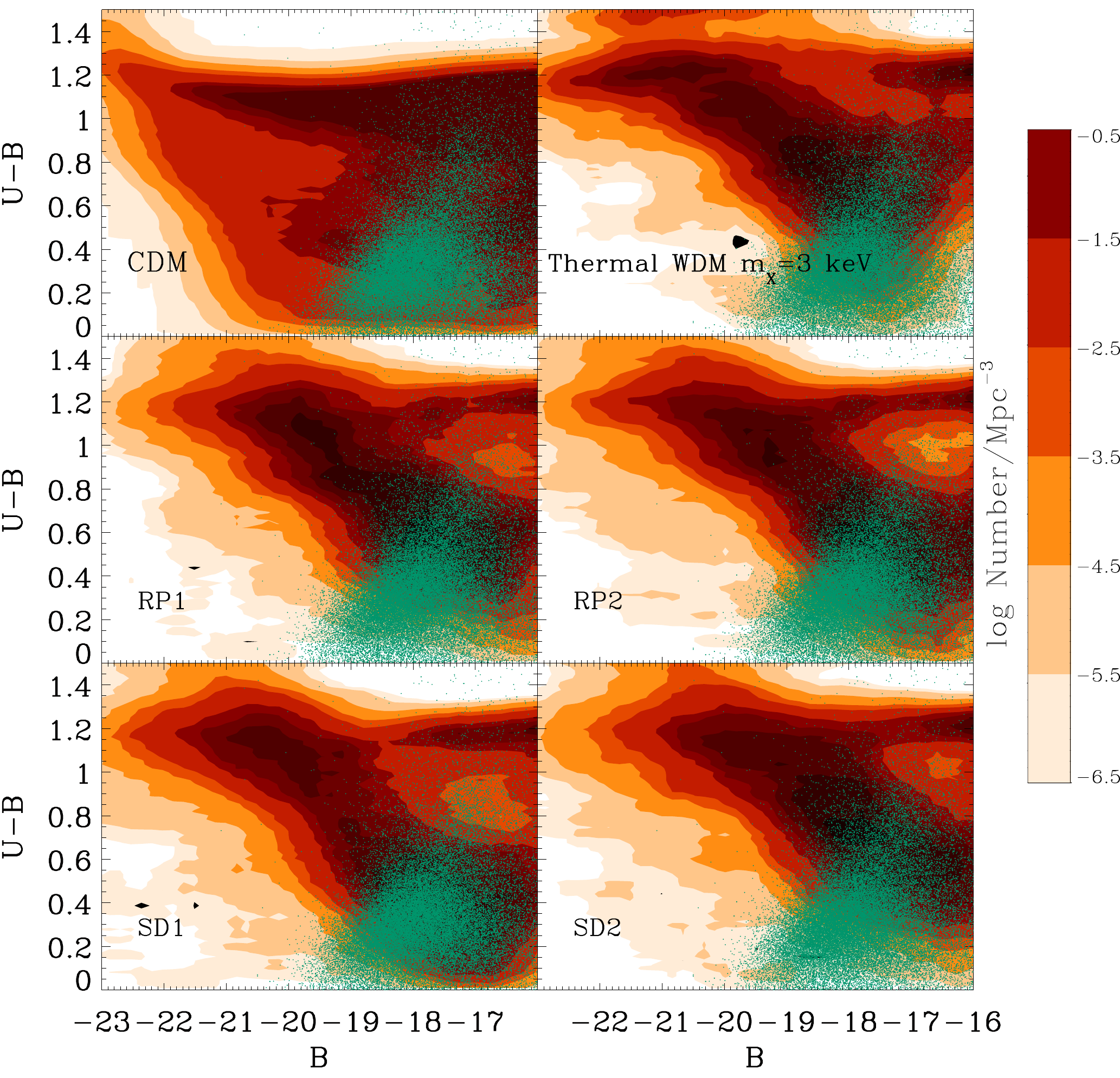}}}
\end{center}
{\footnotesize Figure 11.  For the different DM scenarios considered in the text, we show as a contour plot the distribution of model galaxies for $z\leq 0.1$ in the color-magnitude plane, U-B vs. B (Vega magnitudes).  These are compared with the data from the SDSS for dwarf galaxies with $M_*\leq 10^{9.5}$ $M_{\odot}$ (green points; SDSS magnitudes have been converted using the relations in Fukugita et al. 1996). We included only galaxies brighter than the completeness limit $g=22.2$ of the SDSS survey at $z=0.1$.}
\vspace{0.2cm}

The large fraction of young, star-forming dwarf galaxies obtained
in DM models with suppressed spectra shows up in the distribution
of galaxies in the color-magnitude plane. In fact.  
the U-B color constitutes a proxy for the SSFR of galaxies 
(although it is also affected by dust extinction). 
The bimodality 
in the SSFR distributions discussed above reflects here into 
two classes of galaxies: the ones with bluer colors, mainly contributed
by low-luminosity objects, constituting the ’blue cloud’,
and the ’red sequence’ characterized by red colors $(U-B)\ge 1$, 
mainly contributed by brighter objects. 
The distribution of observed 
points thus defines the position and the relative population of the
dwarf galaxies belonging to the blue cloud.

Although all the DM models yield a bimodal distribution in the color-magnitude plane, 
the WDM and SN models present a more pronounced 
correlation between colour and B-band magnitude 
(see, e. g., Faber et al. 2007; Pierce et al. 2010) compared with the CDM predictions. \\
At the faintest magnitudes ($B\geq -19$) 
the theoretical color distributions computed assuming different DM scenarios 
show the most marked differences. 
In particular, at these magnitudes the CDM model predicts a major fraction of  galaxies to have 
red colours (U-B)$>0.8$, whereas observations indicate that faint galaxies populate preferentially the region 
with (U-B)$\approx 0.3-0.4$ (a long standing problem  
problem of CDM scenarios, see Kimm et al. 2009; Dave' et al. 2011;  Hirschman et al. 2012, Bower et al. 2012, Weinmann et al. 2012; Hirschman et al. 2013). 
On the other hand, in all models with suppressed power spectra  the maximum  density of faint systems is generally 
visible at colours (U-B)$<0.7$, in  better 
agreement with the observational data, which are particularly crowded at 
B$\approx -18$ and (U-B)$\approx 0.3-0.4$. 
Among the considered  models, the one which best reproduces the data is SN SD1, which shows 
a remarkable overlap of the position of its densest region and 
the one of the observational distribution. 

Such a striking difference in the predictions of CDM and the WDM and SN models which concerns the 
abundance of blue, faint galaxies, 
is  due to the combined effects of the strong feedback adopted  in the CDM model (required to comply with the flat slope of the local stellar mass functions)
 and of the delayed star formation histories characterizing the WDM and SN models. 
In the future, in order to gain more clues on the nature of DM and to further probe 
the DM models, more observables will need to be investigated, possibly sensitive to the presence of young 
stellar populations in dwarf galaxies in particular. These observables may include, e.g., the gas-to-stellar  
mass fractions or the cold gas content in general of such systems as well as their star formation history, 
generally addressable by means of color magnitude diagrams (e.g., Vincenzo et al. 2016; Sacchi et al. 2016).

\section{Summary and Conclusions}

We have used a semi-analytic model (SAM) to investigate galaxy formation in cosmological models with dark matter constituted by sterile neutrinos. We focused on models with fixed sterile neutrino mass $m_{\nu}=7$ keV, consistent with the tentative 3.5 keV line recently detected in several X-ray spectra of clusters and galaxies. Specifically, we chose to focus on SN models which are  marginally consistent with existing bounds but still yielding an appreciable suppression of the power spectrum with respect to CDM at scales $M\lesssim 10^{9}\,M_{\odot}$. In particular we considered:
\begin{itemize}
\item two resonant  production models of sterile neutrino with mixing angles $\sin^2(2\theta)=2\,10^{-10}$ (model RP1) and  $\sin^2(2\theta)=5\,10^{-11}$(model RP2), to cover the range of mixing parameter which are consistent with the tentative 3.5 keV line;
 \item two scalar-decay models representative of the two possible cases characterizing such a scenario, a freeze-in (model SD1) and a freeze-out case (model SD2), both with coupling parameter $10^{-8.5}$ between the scalar and the sterile neutrino fields, but with different coupling of the scalar with the Higgs field.
 \item For comparison, we also considered the standard CDM case and the thermal WDM with particle mass $m_X=3$ keV. 
\end{itemize}
Following the approach of previous works on the comparison between CDM and WDM (e.g., Kennedy et al. 2014; Wang et al. 2017), we calibrated the feedback parameter of our SAM to match the shape of the local stellar mass distribution, and investigated the effects of assuming the different DM models on the model predictions 
comparing the result of our SAM to a wide set of observables. The aim is to investigate to what extent the comparison with the different observables can help to disentangle the effects of baryon physics (in particular of feedback) from the specific effects of the different assumed DM models, so as to single out the most promising observational probes for the DM scenario driving galaxy formation. We found that: 
\begin{itemize}
\renewcommand{\labelitemi}{$\bullet$}
\item The  stellar mass function of satellites of Milky Way-like galaxies is prone to the degeneracy between the effects of feedback and those related to the DM power spectrum. These are both effective in yielding  satellites  abundances consistent with recent observations. Nevertheless, the predictions of both the thermal WDM and the SD1 models are in tension with present data since the corresponding predicted abundances are below (at more than 1-$\sigma$ c.l.) the observed data. 
\item  Measurements of the stellar-to-halo mass ratios in low mass galaxies constitute in principle an effective way to disentangle the effects of feedback from those 
related to the DM power spectrum. All the considered DM models yield a large fraction of dwarf galaxies with $-4\leq log (M_*/M)\leq -2$ for $M\leq 10^{9}\,M{\odot}$ at variance with the CDM scenario. However, present data are too sparse (due to the observational biases discussed in sect. 3) to provide a definite evidence for such ratios. 
\item The abundance of faint ($M_{UV}\geq -12.5$) galaxies in the UV luminosity functions at redshifts $z\geq 6$ constitutes at present the most clean way to probe 
 DM scenarios based on WDM or on sterile neutrinos. Even maximizing the  systematics effects affecting present observations (see Bouwens 2017a,b; Kawamata et al. 2017), the SD1 model is excluded at more than 1-$\sigma$ confidence level. Next improvements into the measurements of the faint end of the UV luminosity function will provide a powerful probe for DM models based on SNs. 
\item The star formation properties of dwarf galaxies (stellar masses $M_*\sim 10^{9}\,M_{\odot}$) depend on the assumed DM model; DM models with suppressed power spectra are characterized by a delay in the stellar mass growth history ranging from 500 Myr (RP1 and SD2 model) to $\approx 1$ Gyr. This yields for such models a larger fraction of active (SSFR $\geq 10^{-11}$ yr$^{-1}$) galaxies with blue colors ($U-B\leq 1$), and young age $\leq 10^{9}$ yr compared to the CDM case, providing a better match to present data. Such conclusions are robust with respect to the variation in the feedback efficiency {\it when} the latter is calibrated in each models so as to match the slope of the local stellar mass function. 

\end{itemize}

The comparison with existing works in the literature on galaxy formation in thermal WDM models (and in SN DM models, when comparing with the Milky Way satellites)  supports the sensibleness of our approach and the robustness of our results. Indeed, our strategy for fixing the feedback parameters follows that adopted in recent works based on SAM (see, e.g., Kennedy et al. 2014, Lovell et al. 2016; Wang et al. 2017). For the abundance of satellite galaxies in the thermal WDM case (with $m_X=$ 3 keV) we recover the tension between the predicted and the observed values found in Kennedy et al. (2014) for our assumed value of the mass of Milky Way-like galaxies ($M=1-2.5\cdot\,10^{12}\,M_{\odot}$). When we compare our RP1 and RP2 models with the LA8 and LA12 models explored by Lovell et al. (2017a)  (characterized by similar sterile neutrino mass and mixing parameters) we recover similar results. The larger $M_*/M$ ratios that we find in SN and in the WDM models compared with the CDM case is similar to what found in previous works in the literature (e.g., Guo et al. 2011 for the CDM case, Corasaniti et al. 2017 for the WDM case), while the dramatic increase of the scatter of the $M_*/M$ relation that we find  for decreasing masses $M\lesssim 10^9$ $M_{\odot}$ is consistent with what  found in recent N-body simulations (see Munshi et al. 2017). The small effect of assuming a WDM 
 spectrum on the shape and evolution of the stellar mass function (with a mild increase in the abundance of high-mass galaxies in models with suppressed power spectrum) is consistent with the findings in Wang et al. (2017) for their thermal WDM model, while 
the effects of assuming a WDM power spectrum on the UV luminosity function at high redshift is comparable to what found by Dayal et al. (2015). Finally,   the delay in the growth of the stellar mass in WDM models is quantitatively close to what found by Governato et al (2015). Thus, on the theoretical side, 
 our main conclusions fit into the framework that is being outlined by different groups in overlapping cases. 

On the observational side, next  efforts will soon provide key tools to probe the role of the DM models based on SN in driving galaxy formation. 
In the near future, the wide survey of LSST (Ivezic et al. 2016) will
cover the whole southern sky at a depth which is around 4 magnitudes
deeper than the SDSS: the combination of color
information and excess in the surface density of stars will allow to significantly improve the
detection of dwarf satellites in the Milky Way and to extend it to stellar 
masses $M_*\sim 10^4 M_{\odot}$ at a distance of
$\sim$1Mpc. 
To robustly assess the total masses of these ultra-faint dwarfs,
  a quantum leap forward will be provided by dynamical
measurements with JWST, SKA, and its precursors (ASKAP, MeerKAT, MWA,
and HERA). In particular, spectroscopic IFU observations with JWST on
ultra-faint dwarfs could constrain the dynamical mass of these objects
by determining the stellar velocity dispersion and/or the gas rotation
with nebular lines up to large distances from the dwarf
center. Particular care however should be paid to mitigate the effects
of inclination on the dynamical mass estimates and on the role of the
environment on the observed baryonic to total mass ratio. Similarly,
SKA and its pathfinders could measure the HI velocity curves of faint
galaxies up to large distances from the center, in order to reduce the
systematic biases in the total mass estimates. These dynamical mass
measurements will allow to break the present degeneracies between
baryonic feedback and the nature of DM, degeneracies which can be
still present when comparing the number statistics of dwarf galaxies
(see Fig. 4).

As for the abundance of high-$z$, faint galaxies, our knowledge of the evolution of the UV LF (Fig. 7) will be
significantly improved by JWST observations: NIRCam ultra-deep imaging
at 1-5$\mu$ enables the selection of high-redshift galaxies $>$1
magnitude fainter than in present HST samples (Finkelstein et
al. 2015) extending UV LF estimates both in luminosity and redshift
(in principle up to z$\sim$20). At the same time, JWST spectroscopic
observations will also tighten constraints on the evolution of the
star formation rate density (fig. 7) enabling the measurement of star formation rates from recombination lines
(H$\alpha$ at z$<$6.5 and H$\beta$ at higher redshifts), and of dust
extinction from the Balmer decrement. 

Finally, deep JWST will greatly improve the reliability of the
specific SFR (SSFR), by reducing the associated uncertainties by
approx 0.3 dex of their true value. According to simulations, stellar
masses and SFRs will be recovered within 0.2 dex up to z$\sim$9
(Bisigello et al. 2017) and with higher precision at lower redshifts,
thanks to NIRCam multi-wavelength imaging and NIRSpec spectroscopic
capabilities. With this facility, the observed distribution of SSFR
will be extended towards the range of dwarf galaxies by at least an
order of magnitude in stellar mass.

\begin{acknowledgements}
We would like to acknowledge the financial support of ASI (Agenzia Spaziale Italiana) under contract to INAF: ASI 2014-049-R.0 dedicated to SSDC.
F. C. acknowledges funding from the INAF PRIN-SKA 2017 program 1.05.01.88.04. 
\end{acknowledgements}

\end{document}